\documentclass[12pt]{article}          
\usepackage{amsmath}             
\usepackage{graphicx}            \usepackage{amssymb}
\usepackage{appendix}
\usepackage{slashed}
\usepackage{empheq}
\usepackage{comment}
 \font\tenrm=cmr10

\usepackage{geometry}

\title{Interpretation Neutrality in the Classical Domain of Quantum Theory}  
\author{Joshua Rosaler}                                       
\date{}

\newcommand*\widefbox[1]{\fbox{\hspace{2em}#1\hspace{2em}}}

\begin{document}   








          
\maketitle

\begin{abstract}             
\baselineskip=10pt \tenrm

I show explicitly how concerns about wave function collapse and ontology can be decoupled from the bulk of technical analysis necessary to recover localized, approximately Newtonian trajectories from quantum theory. In doing so, I demonstrate that the account of classical behavior provided by decoherence theory can be straightforwardly tailored to give accounts of classical behavior on multiple interpretations of quantum theory, including the Everett, de Broglie-Bohm and GRW interpretations. I further show that this interpretation-neutral, decoherence-based account conforms to a general view of inter-theoretic reduction in physics that I have elaborated elsewhere, which differs from the oversimplified picture that treats reduction as a matter of simply taking limits. This interpretation-neutral account rests on a general three-pronged strategy for reduction between quantum and classical theories that combines decoherence, an appropriate form of Ehrenfest's Theorem, and a decoherence-compatible mechanism for collapse. It also incorporates a novel argument as to why branch-relative trajectories should be approximately Newtonian, which is based on a little-discussed extension of Ehrenfest's Theorem to open systems, rather than on the more commonly cited but less germane closed-systems version.  In the Conclusion, I briefly suggest how the strategy for quantum-classical reduction described here might be extended to reduction between other classical and quantum theories, including classical and quantum field theory and classical and quantum gravity.

 \end{abstract}

\section{Introduction}

As several authors have noted, a full quantum-mechanical account of classical behavior requires an explanation of both the ``kinematical" and ``dynamical" features of classicality.  
Kinematical features include determinate values for properties such as position and momentum, separability or effective separability of states across different subsystems, and others.
Dynamical features of classicality, on the other hand, consist primarily in the approximate validity of classical equations of motion. 
Most work on quantum-classical relations focuses on recovering one or the other aspect of classicality, but not both together.    
\footnote{The kinematical/dynamical terminology employed here follows Bacciagaluppi's \cite{bacciagaluppi2011measurement}.
}
For example, literature on semi-classical analysis, including discussions of the WKB approximation, $\hbar \rightarrow 0$ limits and various formal quantization procedures, focuses primarily on the recovery of classical equations of motion and related mathematical structures of classical mechanics but tends to sidestep questions about the recovery of determinate measurement outcomes and other kinematical features of classical behavior. On the other hand, most literature on quantum measurement focuses on kinematical features of classicality while paying relatively little attention to the problem of explaining the approximate validity of Newton's equations. 

However, an important subset of the literature on decoherence theory goes further than most sources toward recovering both the kinematical and dynamical features of classicality in a unified way. Predictably, though, where the recovery of kinematical features is concerned, these analyses come up against the notorious difficulties associated with collapse of the quantum state. 
\footnote{I will take the term ``collapse" here to encompass both real, dynamical collapse processes as well as processes in which collapse of the quantum state is merely effective or apparent. 
}
While decoherence offers a promising mechanism for suppressing intereference among, and in some sense defining, the various possible ``outcomes" represented by the quantum state, it does not by itself explain why only one of these alternatives appears to be realized, much less with the specific probabilities given by the Born Rule. Further explanation, which goes beyond the resources furnished by decoherence theory, is required to account for the phenomenology of Born Rule collapse. 
The present investigation shows how results from decoherence theory can be combined with the specific collapse mechanisms and ontologies associated with different realist interpretations of quantum theory in order to give a more complete - if also more speculative - account of both the kinematical and dynamical features of classical behavior. While the strategy for quantum-classical reduction summarized here reflects an understanding of classical behavior that is implicitly held - at least in its broad outlines - by many experts on decoherence theory, I seek to bring this picture into sharper focus by consolidating insights that remain dispersed across the literature and by making explicit several points that have not been sufficiently emphasized or developed. My analysis here aims in particular to bring out two important features of the decoherence-based framework for recovering classical behavior:   


\vspace{5mm}

\noindent \textbf{1) Interpretation Neutrality} 

I offer a detailed development of the view that the bulk of technical analysis needed to recover classical behavior from quantum theory is largely independent of the precise features of the collapse mechanism and ontology of the quantum state. The analysis below demonstrates that concerns about wave function collapse and ontology can be addressed as a coda - allbeit a necessary one - to the account of classical behavior suggested by decoherence theory, so that one need not start anew in the recovery of classical behavior with each new interpretation
\footnote{As several authors have noted, different ``interpretations" of quantum mechanics, such as the Everett, de Broglie-Bohm and GRW interpretations, are more properly regarded as separate \textit{theories} since they differ in the accounts of physical reality (in particular, the laws and ontology) that they take to underwrite the success of the quantum formalism. Nevertheless, I will conform to common usage in referring to them as ``interpretations" of quantum theory. 
}
of quantum theory that is considered. The interpretation-neutral strategy for recovering classical behavior summarized here rests on three central pillars: decoherence, which generates a branching structure from the unitary quantum state evolution, such that the state of the system of interest relative to each branch is well-localized; Ehrenfest's Theorem for open quantum systems, which ensures that the only branches with non-negligible weight are branches relative to which the system's trajectory is approximately Newtonian; a decoherence-compatible prescription for collapse, which serves at each moment to select just one of the branches defined by decoherence in accordance with the Born Rule. The underlying physical mechanism (associated with some particular interpretation of quantum mechanics) for this decoherence-compatible collapse is left unspecified. 

This analysis counters the notion promulgated by some authors that recovering classical behavior from quantum theory is a highly interpretation-dependent affair. 
For example, many advocates of the de Broglie-Bohm (dBB) interpretation have defended an approach built entirely around a condition that is more or less unique to dBB theory: namely, the requirement  that the ``quantum potential" or ``quantum force," which generates deviations of Bohmian trajectories from Newtonian ones, go to zero. 
\footnote{It is possible to define the quantum potential and force in other interpretions, but because these other interpretations do not possess localized trajectories at a fundamental level, the quantum potential and force lack the immediate and obvious significance that they posses in dBB theory.
}
Elsewhere, I have argued that the quantum potential is something of a red herring and that the most transparent route to recovering classicality in dBB theory relies primarily on structures common to many interpretations, which are associated with decoherence theory  \cite{RosalerDBB2015}. The present article extends this analysis of classical behavior beyond the context of dBB theory to consider other realist interpretations as well, including the Everett and GRW interpretations. It also provides a more detailed elaboration of the interpretation-neutral, decoherence-based framework for recovering classical behavior.

\vspace{5mm}

\noindent \textbf{2) A Specific Sense of ``Reduction"} 

Second, I show that the interpretation-neutral framework for recovering classical behavior provided by decoherence theory fits a more general model-based picture of inter-theoretic reduction in physics that I have elaborated and defended elsewhere, according to which reduction between theories is based on a more fundamental concept of reduction between two models of a single, fixed system \cite{RosalerLocRed}. Here, I understand a ``model" to be specified by some choice of mathematical state space (e.g., phase space, Hilbert space) and some additional structures on that space that serve to constrain the behavior of the state (e.g., Hamilton's equations, Schrodinger's equation). 
This approach differs in important respects from the more conventional approach to reduction in physics that seeks to recover one theory simply as a mathematical limit of another - typically, in the case of quantum-classical relations, by taking the limit $\hbar \rightarrow 0$ or $N \rightarrow \infty$ -  while recognizing that limits still carry a strong relevance for inter-theory relations in physics. It also differs from approaches to reduction that have been emphasized in the philosophical literature, which aspire to give a completely general account of reduction across the sciences and so fail to capture the strongly mathematical character of reductions within physics specifically. 
One feature that distinguishes the view of reduction employed here from these other approaches is that, rather than attempting to give criteria for reduction directly between \textit{theories} as these other approaches do, it is grounded in a more fundamental and more local concept of reduction between two \textit{models} of a single physical system. Moreover, reduction between models of a single system on this approach is an empirical, \textit{a posteriori} relation between models rather than a formal, \textit{a priori} relation that can be assessed on purely logical or mathematical grounds. While this account of inter-theoretic reduction incorporates insights about reduction previously highlighted by other accounts, its novelty lies in the particular combination of features that it possesses: namely, that it is model-based rather than theory-based, ``local" rather than ``global", and \textit{a posteriori} rather than \textit{a priori}. By highlighting an important sense of reduction and showing how decoherence theory provides a viable framework for effecting this kind of reduction between quantum and classical theories, I seek to provide a counterweight to recent discussions - in particular, by Batterman, Berry and Bokulich - that have urged a move away from thinking about quantum-classical relations as an instance of reduction  \cite{BattermanDD}, \cite{BerryAsymptotics}, \cite{bokulich2008reexamining}. In a separate article, I argue that the singular mathematical limits that Batterman and Berry take to block reduction between classical and quantum mechanics do not block reduction between these theories in the particular sense described here \cite{rosaler2015formal}. 

\vspace{5mm}

Beyond its defense of these two points, much of the novelty of the present discussion lies in its explicit synthesis of various elements from different parts of the literature on decoherence, the measurement problem and the quantum-classical correspondence, and in its attentiveness to nuances that arise when joining these various elements - from the explicit requirement that an interpretation-neutral collapse prescription be decoherence-compatible, to subtle variations in the decoherence conditions required for effective collapse across different interpretations, to the implementation of the open-systems form of Ehrenfest's Theorem rather than the more commonly discussed but less appropriate closed-system form. 
The discussion is structured as follows. Section \ref{Accounts} provides an overview of sources in the decoherence literature that attempt to explain the validity of classical equations of motion and highlights points on which the present discussion serves to complement these investigations. Section \ref{Preliminaries} discusses several important points of terminology and methodology, including my usage of the term ``classical" and a description of the general approach to reduction adopted here. Section \ref{FAPPClassical} describes a framework for recovering classical behavior that is based in results from decoherence theory but that remains non-commital as to the precise mechanism for collapse and the ontology of the quantum state. 
Section \ref{InterpretationDependence} shows how the interpretation-neutral account of classical behavior provided by decoherence can be specially tailored to give an account of classicality in the Everett, de Broglie-Bohm and GRW interpretations, thereby resolving - if only in a speculative way - ambiguities associated with collapse in the interpretation-neutral account. 
Section \ref{DecoherenceConcerns} acknowledges a number of general concerns about decoherence-based approaches to recovering classical behavior. 
Section \ref{Conclusion}, the Conclusion, briefly considers some broader implications of the analysis given here, including possible extensions of this strategy to the relations between quantum and classical field theory and classical and quantum gravity. The core claims of the article are defended in Sections 3, 4 and 5.

\section{Existing Decoherence-Based Accounts of Newtonian Behavior} \label{Accounts}

In this section, I review several important sources in the decoherence literature that attempt to explain the approximate validity of classical equations of motion on a quantum-mechanical basis,
highlighting the respects in which our discussion here differs from, and serves to complement, these other investigations. Generally speaking, none of these sources formulates the decoherence-based account of classical behavior in a way that can be transparently and straightforwardly tailored to accommodate multiple interpretation-dependent collapse mechanisms, and none
explicitly frames its analysis of classicality within the more general account of inter-theoretic reduction in physics that I advocate here.  
 
 In \cite{bacciagaluppi2011measurement} and \cite{sep-qm-decoherence}, Bacciagaluppi describes in verbal, qualitative terms the general approach to recovering classical behavior that I elaborate in more technical detail below.  In particular, he regards decoherence theory as an interpretation-neutral body of results that serves to define the range of possible measurement 
\footnote{I understand ``measurement" here in a generalized sense that takes measurement to be any physical interaction that establishes the appropriate sort of correlation between the system of interest and degrees of freedom external to that system. Crucially, a measurement in this sense need not involve an ``apparatus" or ``observer". 
}
 outcomes associated with different branches of the quantum state, and interpretation-specific collapse mechanisms as serving to select one of these branches in accordance with the Born Rule. 
 \footnote{As Bacciagaluppi observes in \cite{sep-qm-decoherence}, and as I discuss further in Section \ref{InterpretationDependence}, it is an open question whether the dynamical collapse of GRW theory does in fact produce one of the set of states pre-defined by decoherence. 
}
Beyond developing this general approach in more technical detail, our analysis differs from Bacciagaluppi's with regard to the particular form of Ehrenfest's Theorem that it invokes to explain why branch-relative trajectories should be approximately Newtonian.  While Bacciagaluppi invokes the familiar closed-systems version of Ehrenfest's Theorem, our analysis relies on a more germane but less familiar version of this theorem derived by Joos and Zeh for the special class of open systems whose reduced state obeys the Caldeira-Leggett master equation (see \cite{JoosZehBook}, Ch.3). Our analysis carries Joos and Zeh's result further by showing that on timescales where ensemble wave packet spreading of the system in question is sufficiently small, the localized, branch-relative state of the system will approximately follow a Newtonian trajectory, deviating from Newtonian form only through small, stochastic, branch-dependent ``quantum fluctuations" (see Figure ~\ref{fig:NewtonBranching}).

Zurek has shown how classical equations of motion for probability distributions over phase space can be recovered in open quantum systems undergoing decoherence \cite{zurek1998decoherence}. In particular, he has argued that as a result of decoherence, the Wigner function representation of the state of an open quantum system approximately obeys a classical Fokker-Planck equation, which takes the form of a classical Liouvillle evolution combined with a momentum diffusion term. In cases where the corresponding diffusion coefficient is sufficiently small, the evolution of the Wigner function distribution approximates the ordinary classical Liouville evolution. Concerning the more fine-grained level of description in terms of individual trajectories rather than distributions over phase space, Bhattacharya, Habib and Jacobs and Habib, Shizume and Zurek both note that the Fokker-Planck evolution is equivalent to a Langevin equation driven by Gaussian white noise \cite{bhattacharya2000continuous}, \cite{habib1998decoherence}. The diffusion term that contributes to the distribution dynamics reflects the influence of the additional noise term at the level of the trajectory dynamics. 
Rather than motivating the introduction of localized, stochastic quantum trajectories indirectly through the analysis of phase space probability distributions as these authors do, our discussion extracts these trajectories directly from an analysis of the branching structure of the quantum state. This strategy much more readily facilitates the tailoring of decoherence-based results to a variety of different collapse mechanisms and ontologies associated with different interpretations.

Gell-Mann and Hartle approach the recovery of classical equations of motion from the perspective of the decoherent histories framework and the closely related path integral formalism for open quantum systems \cite{gell1993classical}, \cite{hartle2011quasiclassical}. They show that the only trajectories of an open quantum system that contribute non-negligibly to its path integral are approximately Newtonian in form. These results are also discussed in the work of Halliwell \cite{halliwell1995review}. As in the preceding analyses, the roles of branching and collapse in the recovery of determinate Newtonian trajectories are not made transparent in these discussions, nor is the signifcance of these results across different interpretations. 
\footnote{It is worth noting here that Gell-Mann and Hartle regard the formalism of decoherent histories as furnishing its own distinct interpretation of quantum theory. Here, we treat this formalism simply as a useful and elegant framework for analyzing the branching structure of a unitarily evolving state. 
}


It is also worth briefly mentioning several other sources that concern the recovery of approximately Newtonian trajectories within decoherence theory. In \cite{schlosshauer2008decoherence}, Schlosshauer argues that wave packet trajectories should be approximately Newtonian for a narrowly defined set of models and initial conditions, but does not discuss the more general mechanisms and results that give rise to approximately Newtonian trajectories over a wide range of different models and initial conditions, as we do here. In \cite{schlosshauer2005decoherence}, Schlosshauer examines the relationship between decoherence and various interpretations of quantum theory; our discussion in Section \ref{InterpretationDependence} below complements Schlosshauer's by emphasizing that quantitatively different forms of decoherence are needed to induce effective collapse across different interpretations. Landsman's \cite{landsmanClassicalQuantum} cites a special role for decoherence in explaining the dynamical origin of coherent states, which he claims are needed to ensure certain correspondences in the mathematical limit $\hbar \rightarrow 0$.  However, Landsman's focus is primarily on mathematical correspondences between the formalisms of quantum and classical mechanics, rather than directly on the quantum mechanical description of classical \textit{behavior}. 

\section{Terminology and Methodology} \label{Preliminaries}

In this section, I clarify several points of terminology and methodology employed in the discussion below. First, I explain my usage of terms such as ``classical" and ``quasi-classical". Second, I describe a model-based sense of reduction in physics that I have elaborated in detail elsewhere; in later sections, I argue that decoherence theory provides a viable strategy for effecting this type of reduction between quantum and classical mechanics. 


\subsection{The Meaning of ``Classical"}

The term ``classical" carries a range of different meanings across the literature on quantum-classical relations. In the context of recovering classical behavior within quantum theory, I take it to denote the following kinematical and dynamical requirements: 

\

\underbar{Kinematical Requirements}
\begin{itemize}
\item \textit{quasi-classicality:} The system in question must possess determinate or approximately determinate 
values for position and momentum simultaneously; within the constraints of the uncertainty principle, it is possible for both quantities to be simultaneously sharply defined relative to macroscopic scales of length and momentum. 
\item \textit{separability:} One some level, the state of a system exhibiting classical behavior should be determined completely by the states of its individual subsystems.
\footnote{Of course, in quantum theory this is not the case for general states since the density matrix of two subsystems does not generally determine the density matrix of the combination of these two systems. 
}
\item \textit{``objectivity":} Disjoint subsystems of the environment - including those associated with any observers that may be present -  must agree on the quasi-classical values for position, momentum or other variables that they register for the system of interest. This requires a strong measure of redundancy in the encoding of information about the central system within the environment (see, e.g., \cite{zurek2003quantum}, \cite{schlosshauer2008decoherence}, Section 2.9 and references therein for further elaboration of this concept of objectivity; see  \cite{fields2014ollivier} for a critique of this notion).
\end{itemize}

\underbar{Dynamical Requirements}
\begin{itemize}
\item \textit{obeying classical equations:}  Quasi-classical trajectories must approximate the solutions to classical equations of motion over appropriate timescales and within appropriate margins of error. As I discuss below, these timescales and margins of error are constrained by the timescales and margins of error within which the classical model of a system successfully tracks the system's behavior. 
\end{itemize}

\noindent Examples of the sort of system whose classical behavior we seek to model quantum mechanically include the following:

\

\begin{itemize}
\item center of mass of the Moon, planets, etc. 
\item center of mass of a golf ball, baseball, etc.
\item charged particle (e.g., alpha particle, proton, electron) in a bubble chamber or particle accelerator.
\end{itemize}

\noindent It is important to note, as others have, that the set of systems that are well-described by purely classical models includes not only macroscopic bodies but also (as in the last set of examples) bodies of atomic and sub-atomic size. For example, in a particle accelerator or bubble chamber, it is classical equations - specifically, the Lorentz force law - that are used to predict and guide the particles' motion. 
It is also worth taking a moment to note that the term ``classical" is sometimes used to describe phenomena for which a certain \textit{hybrid} of quantum and classical concepts is effective - for example, in studies of semi-classical analysis and applications of the WKB approximation in quantum physics (see, for example, \cite{BattermanDD}, Ch.7 and \cite{BerryAsymptotics}). I do not employ the term in this manner, reserving it instead for behavior that is well-described by models that are \textit{exclusively} classical and do not contain any reference to quantum mechanical concepts.

\subsection{A Local, Empirical, Model-Based Concept of Inter-Theoretic Reduction in Physics} \label{InterModelReduction}

The term ``reduction" is notoriously slippery. In the context of inter-theory relations in physics, the term is often associated with the requirement that one theory be a ``limit" or ``limiting case" of another. However, as I have argued in detail elsewhere, this requirement is extremely vague since the limit of a theory is not a mathematically well-defined concept (nor, for that matter, is the concept of a theory) and it is far from clear what it means, in general, for one theory to be a limit of another \cite{RosalerLocRed}. To judge from common occurrences of this manner of speaking,
it seems that the limit-based concept of reduction is defined generally by little more than the very weak requirement that somehow, something like a limit be involved in the relationship between some portions of the two theories involved. 

Within the philosophical literature on inter-theoretic reduction across the sciences, the term ``reduction" is sometimes associated with an account of the concept proposed by Ernest Nagel and later refined by Kenneth Schaffner. Schaffner's more widely accepted refinement of Nagel's account requires that it be possible to deduce approximations to the laws of one theory from those of another using propositions known as ``bridge principles" that relate terms in the high-level (i.e., less encompassing) theory that do not occur in the low-level (i.e., more encompassing) theory to terms in the low-level theory. However, there is no established view among advocates of this approach as to what the general nature of these bridge principles is - e.g., whether they are defintions or empriically established claims, biconditional identities or one-way conditional associations, linguistic propositions or something else. 
An alternative to the Nagel/Schaffner account of that is occasionally discussed in the philosophical literature is the model of Kemeny and Oppenheim, which was also intended to apply generally to reduction across the sciences. Briefly summarized, Kemeny and Oppenheim's account requires that the reducing (more encompassing) theory explain all of the observational data explained by the reduced (less encompassing) theory and that the reducing theory be at least as ``well systematized" as the reduced theory. 
Roughly, the degree of systematization of a theory is understood as the ``ratio" of the amount of observational data explained by the theory to the number of fundamental assumptions that serve to define the theory. One difficulty for the Kemeny-Oppenheim account of reduction has been to make this concept of systematization precise, while another has been its heavy reliance on the observational-theoretical distinction, which has been widely discredited since the decline of logical positivism and empiricism. Because they aspire to give accounts of reduction across the sciences and not just in physics, neither the Nagel/Schaffner nor the Kemeny/Oppenheim model of reduction fully captures the specifically mathematical character of reductions within physics. 

Elsewhere, I have argued for an alternative model-based concept of reduction in physics that incorporates important insights from each of these accounts while correcting some of their shortcomings \cite{RosalerLocRed}. Rather than defining reduction in physics as a matter of effecting a single wholesale derivation of one theory's laws from those of another, as the Nagelian and limit-based accounts tend to do, the account of reduction in physics that I propose defines reduction between theories in terms of a more fundamental notion of reduction between two models that represent the same physical system - for example, between the quantum mechanical and quantum field theoretic models of a single electron, or between the classical and quantum models of an everyday object like a baseball. The specification of a model typically consists in the specification of a state space and some dynamical or other constraints on the behavior of these states. For example, in models of quantum mechanics, this entails a commitment to a particular Hilbert space of some specified dimension
(that is, assuming a non-algebraic approach to quantum theory) with some particular Hamiltonian defined over it.  In models of classical mechanics, it typically entails a commitment to a particular phase space or configuration space of fixed dimension and topology, also with some particular Hamiltonian or Lagrangian. Thus, following van Fraassen and other advocates of the ``semantic" view of theories, a theory can be associated with some category of models - for example, in the case of quantum mechanics, the category of models that can be formulated in terms of the unitary evolution of some Hilbert space vector - while a choice of particular model within that category to describe some particular system involves more detailed specifications  (e.g., of a specific Hilbert space and Hamiltonian) \cite{van1980scientific}. 

Specifically, I take ``theory $T_{h}$ reduces to theory $T_{l}$" to mean that every circumstance under which a real system's behavior is accurately represented by some model of theory $T_{h}$ is also a circumstance in which that same behavior is represented at least as accurately, and in at least as fine-grained a way, by some model of theory $T_{l}$.
\footnote{The subscripts ``h" and ``l" stand for ``high-level" and ``low-level", respectively.
}
Thus the primary requirement for ``reduction" as I use the term here is that the domain of applicability of the reduced theory be subsumed into that of the reducing theory. Because every system in the domain of a theory is represented by some particular model of that theory, this notion of reduction between theories can be understood in terms of the more fundamental notion of reduction between two models of a single system, which I designate $reduction_{M}$. Reduction between two theories, $reduction_{T}$, can be defined in terms of reduction between two models of a single physical system, $reduction_{M}$, as follows:

\begin{quote}
Theory $T_{h}$ $reduces_{T}$ to theory $T_{l}$ iff for every system $S$ in the domain of $T_{h}$ - that is, for every physical system $S$ whose behavior is accurately represented by some model $M_{h}$ of $T_{h}$ - there exists a model $M_{l}$ of $T_{l}$ also representing $S$ such that $M_{h}$ $reduces_{M}$ to $M_{l}$.
\end{quote}

\noindent $Reduction_{M}$ is a three-place relation between $M_{h}$, $M_{l}$ and some real physical system $S$
\footnote{There are at least two ways of clarifying the meaning of the term ``physical system". The first is theory-dependent. For example, while classical mechanics associates the concept of a system with a set of degrees of freedom represented in some configuration space, quantum theory associates it with a set of degrees of freedom represented in some tensor product Hilbert space (and the de Broglie-Bohm interpretation employs both Hilbert space and configuration space). The second sense of the term ``system" seeks to be independent of any particular theoretical framework. It reflects the sense of ``systemÓ in which, for example, the models of an electron in classical mechanics, nonrelativistic quantum mechanics, relativistic quantum mechanics and quantum field theory all represent the same ``physical system" consisting of some particular electron in nature. Admittedly, this second sense of ``system" is more difficult to pin down since it doesn't have the benefit of a fixed theoretical framework to rely on. Nevertheless, it is a concept that we cannot do without, since it is frequently necessary to speak of a particular system without committing to any single theoretical representation of that system, as when one considers relations between distinct representations of the system. The requirements of $reduction_{M}$ invoke the second sense of ``system" when referring to $S$.
}
, which requires that $M_{l}$ provide at least as accurate and detailed a description of $S$'s behavior as $M_{h}$ in all cases where $M_{h}$ is successful. 
Where both models successfully describe (to within some margin of approximation) the same features of $S$, consistency requires a certain dovetailing between the models, which is typically manifested as a certain mathematical relationship between the models. However, it is important to stress that $reduction_{M}$ \textit{only} requires this dovetailing to occur in cases where $M_{h}$ furnishes an accurate representation of $S$, and that the models are permitted to diverge arbitrarily outside of this domain. The particular form of the mathematical relationship between $M_{h}$ and $M_{l}$ that underwrites $reduction_{M}$ in a given case will depend on the general classes of model to which $M_{h}$ and $M_{l}$ belong. Moreover, this relationship will typically be parametrized by empirical values characterizing the accuracy and scope of $M_{h}$ in its representation of $S$. In  \cite{RosalerLocRed}, I give precise conditions for $reduction_{M}$ in a class of cases where both models are deterministic dynamical systems, and briefly suggest how these criteria might be generalized. Below, I specify conditions for $reduction_{M}$ in the class of cases relevant to us here, concerning the reduction of a deterministic dynamical systems model to a stochastic dynamical model. 

It is worth highlighting two features of this model-based approach to reduction in physics. First, it is ``local" in the sense that it allows reduction between theories to be effected through numerous distinct, context-specific inter-model derivations across different systems or classes of system, rather than requiring the existence of a single ``global" derivation that applies uniformly across the entire domain of the high-level theory. 
Second, this kind of reduction is that it is an \textit{a posteriori}, or ``empirical", rather than an \textit{a priori}, or ``formal", relationship between theories/models. While it is often supposed that reduction in physics is solely a feature of the mathematical or logical relationship between two theories or models, the question of whether one representation succeeds at describing the world in all cases where the other does often has an unavoidable empirical component.  The distinction between formal and empirical approaches to quantum-classical reduction is discussed further in 
\cite{rosaler2015formal}.

In the context of the reduction between classical and quantum models of a single system, the classical model can be formulated as a determinsitic dynamical system while the precise nature of the quantum model is interpretation-dependent - for example, models of the Everett and dBB interpretations are deterministic while models of GRW theory are stochastic. However, as I explain in detail below, by relying on the branching structure that arises through decoherence, it is possible to define an effective, stochastic model of the branch-relative quantum state evolution that applies across many different interpretations of quantum theory. 
\footnote{Such effective stochastic quantum evolutions can be modeled using the formalism of quantum stochastic calculus; see, for example, \cite{ford1988quantum} and \cite{ford1987quantum} for details.
}
This effective model assumes a decoherence-compatible type of collapse, but leaves the detailed physical mechanism governing collapse unspecified. What this physical mechanism happens to be varies according to interpretation. Thus, while the reduction of the deterministic classical model to the effective, stochastic quantum model of a system is an interpretation-neutral affair, the task of underwriting the latter model's collapse process within a particular interpretation-specific model is not. 

\vspace{5mm}

The reduction of the classical model to the effective, stochastic quantum model of a system $S$ falls into the category of inter-model reductions in which a deterministic dynamical systems model is reduced to a stochastic dynamical model sharing the same time parameter. It is possible to state formal requirements for  $reduction_{M}$ in this class of cases that are quite general. Denote the state space of $M_{h}$ by $S_{h}$ and the unique time-$t$ evolution of the initial state $x_{h} \in S_{h}$ by $D_{h}(x_{h},t)$. Denote the state space of $M_{l}$ by $S_{l}$ and a stochastic time-$t$ evolution of the initial state $x_{l} \in S_{l}$ by $\mathcal{D}_{l}(x_{l},t)$. Note that $\mathcal{D}_{l}(x_{l},t)$ is not a function in the strict mathematical sense since the time-$t$ evolution of a given initial state does not have a single predetermined value, but emerges probabilistically. Let $d_{h}$ be the set of initial conditions $x_{h}$ in $S_{h}$ such that $D_{h}(x_{h},t)$ tracks the evolution of the system $S$ within margin of error $\delta$ over a time greater than or equal to $\tau$. While there is some flexibility in the choice of values for $d_{h}$, $\delta$ and $\tau$, bounds on these values are empirically constrained by the quality and scope of fit between $M_{h}$ and $S$. 
Given a triple $(\delta, \tau, d_{h})$, reduction between the models requires that there exist a time-independent function $B(x_{l})$ 
\footnote{Note that ``time-independent" here means that the function $B$ is not \textit{explicitly} time-dependent; it may depend indirectly on time through any time dependence of $x_{l}$.
}
from the state space $S_{l}$ of the low-level model to the state space $S_{h}$ of the high-level model and some set $d_{l}$ of states in $S_{l}$ such that $B(d_{l}) = d_{h}$ and for all $x_{l} \in d_{l}$, 

 \begin{empheq}[box=\widefbox]{align} \label{ComRed1}
& \bigg| B(\mathcal{D}_{l}(x_{l},t)) - D_{h}(B(x_{l}),t)  \bigg| < 2 \delta,
\end{empheq}

\noindent or less formally,

\begin{equation} \label{ComRed}
B(\mathcal{D}_{l}(x_{l},t)) \approx D_{h}(B(x_{l}),t), 
\end{equation}

\noindent for all $0 \leq t < \tau$, with probability $1- \epsilon$ for very small $\epsilon$. 
\footnote{The margin of error $\delta$ is left implicit in the approximate equality of equation (\ref{ComRed}).
} 
The factor of $2$ in $2 \delta$ is included because, if the high-level trajectory approximates the system's evolution to within margin $\delta$ and the trajectory induced by the low-level dynamics through $B$ also approximates the system's evolution to within $\delta$, as $reduction_{M}$ requires, then these trajectories may be separated from each other by as much as $2 \delta$. 

\begin{figure}[t]
\includegraphics[scale=0.25]{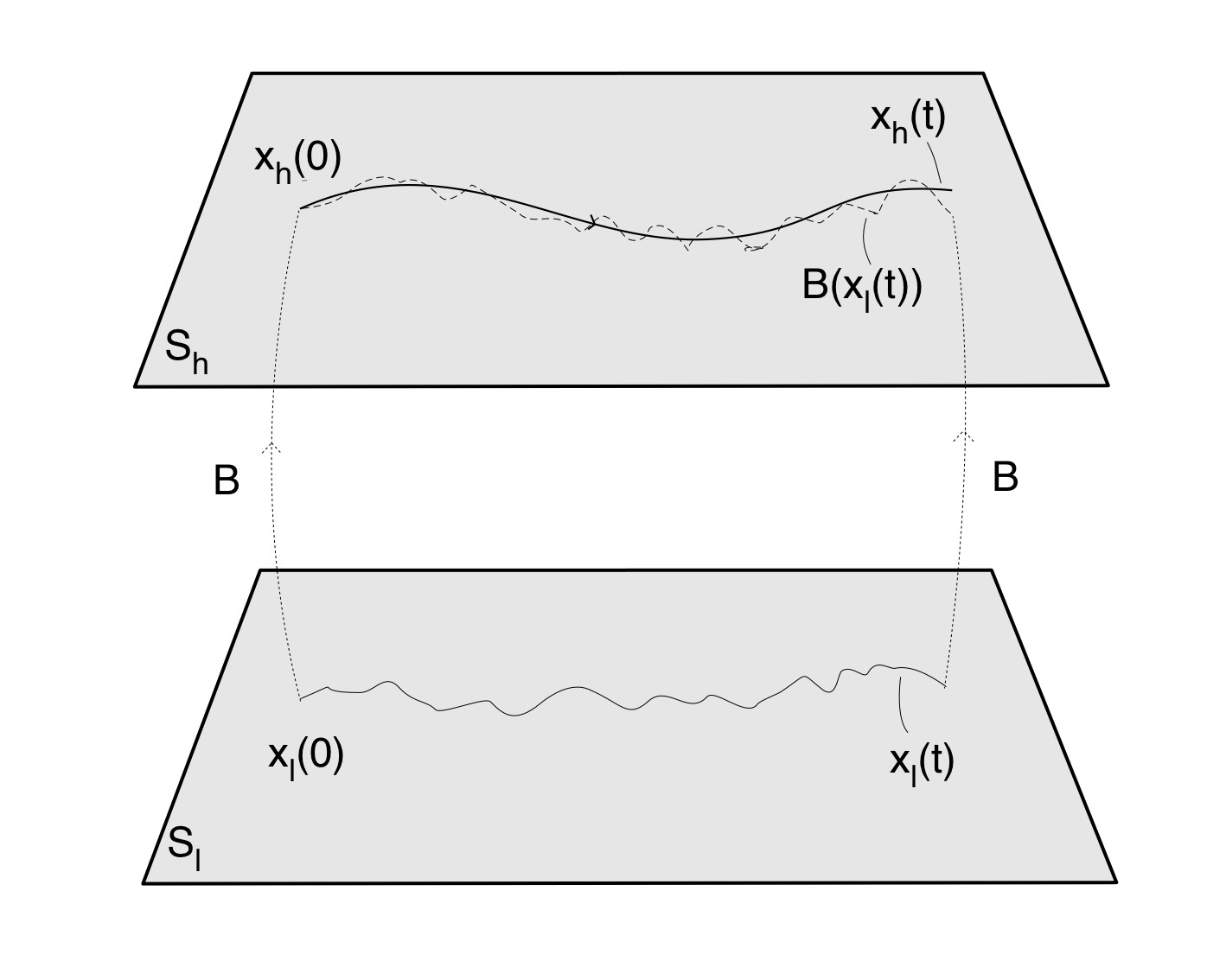}
\caption{\footnotesize In cases where the high-level model of a system is deterministic, the low-level model is stochastic and both models share a time parameter, 
$reduction_{M}$ requires that the trajectories on $S_{h}$ induced via $B$ by the low-level dynamics approximate those high-level trajectories that successfully track the behavior of the physical system $S$. 
}
\label{fig:BridgeMapDiagram}
\end{figure}

Briefly summarized, $reduction_{M}$ in this class of cases requires that with very high likelihood, the operations of dynamical evolution and application of the ``bridge map" or ``bridge function" $B$ approximately \textit{commute} within the relevant empirically determined margins. Equivalently, $reduction_{M}$ requires that all physically salient high-level trajectories be approximated within appropriate margins by trajectories that are \textit{induced} on the high-level state space by the low-level dynamics through the bridge map (see Figure ~\ref{fig:BridgeMapDiagram}). From a physical point of view, this requirement may be interpreted as follows. 
Since $reduction_{M}$ requires that $M_{l}$ succeed at describing $S$'s behavior in all cases where $M_{h}$ does,
when the evolution of $x_{h}$ successfully tracks real features of $S$, 
there must be some quantity defined within $M_{l}$ that, with very high likelihood, tracks these same features at least as accurately. The quantity $B(x_{l})$, which is the same type of mathematical object as $x_{h}$, but whose dynamics are determined by $M_{l}$, fills this role if the requirement (\ref{ComRed}) for reduction holds. Thus, the requirement (\ref{ComRed}) helps to ensure that $B(x_{l})$ provides an adequate surrogate for $x_{h}$ in cases where $M_{h}$ is successful at describing $S$'s behavior. The possibility of using both deterministic and stochastic dynamical models simultaneously to represent the same physical system, as we do here, has been analyzed in detail by Werndl \cite{Werndl1}, \cite{Werndl2}, \cite{Werndl3}, \cite{Werndl4}. 

\subsection{A Few Caveats}

The local, model-based, empirical approach to reduction adopted here suggests a precise mathematical criterion for reduction between classical and quantum models of a single system exhibiting classical behavior. These criteria, which follow the general pattern of eq. (\ref{ComRed1}), are formulated in Section \ref{NewtonianTrajectories}. However, no attempt is made here to provide a rigorous proof that this condition holds across all systems where classical behavior occurs. It is likely that many of the details required to give such a proof will vary across different systems, so that full rigor can only be achieved through specialization to particular systems or classes of system. My goal here is only to outline the central \textit{strategy} offered by decoherence theory for explaining why this condition should hold generally across systems that exhibit classical behavior - that is, to describe a general template into which a more complete derivation could be fit.

\section{An Interpretation-Neutral, Decoherence-Based Strategy for Quantum-Classical Reduction} \label{FAPPClassical}
 
In this section, I outline an interpretation-neutral, decoherence-based strategy for recovering classical behavior from quantum theory, which relies heavily on the general approach to reduction outlined in the previous section. In a nutshell, this strategy rests on three central pillars: 1) decoherence to generate a branching structure for the quantum state, such that the state of the system of interest relative to any single branch is always quasi-classical; 2) a decoherence-compatible collapse prescription that selects one of these branches in accordance with the Born Rule, and whose underlying physical mechanism is left unspecified; 3) an appropriate form of Ehrenfest's Theorem to ensure that quasi-classical, branch-relative trajectories are approximately Newtonian over relevant timescales. In Section \ref{InterpretationDependence}, I show how concerns about the physical mechanism underpinning this decoherence-compatible collapse prescription can be addressed as a coda - albeit a necessary one - to the interpretation-neutral account given in this section, and that interpretation-specific aspects of the quantum mechanical description of classical behavior can therefore be quarantined to a relatively narrow portion of the overall analysis. In other words, we will see how the different collapse mechanisms and ontologies associated with different interpretations can be ``slotted in" to this interpretation-neutral account to give a more complete, if more speculative, picture of classical behavior than the one provided by the interpretation-neutral picture alone. Thus, we will see more explicitly than previous investigations how one can go quite far in providing a quantum-mechanical account of classical behavior without taking on the speculative metaphysical commitments associated with some particular interpretation of quantum theory. Of course, we must also keep in mind that at most one of these interpretation-specific accounts can be correct as a description of the mechanism that nature itself employs.  

We begin by reviewing the basic setting common to decoherence-based accounts of classical behavior, progressively buiding toward the recovery of localized Newtonian trajectories. As we will see, environmental decoherence relative to a pointer basis of coherent states implies decoherence of histories defined with respect to an approximate PVM on phase space. To each such history is associated a branch vector and a corresponding localized (though not necessarily Newtonian) trajectory through classical phase space. Using these branch vectors, it is possible to define an effective, decoherence-compatible prescription for collapse that serves to ``black-box" all interpretation-specific details within a single portion of the analysis - where by ``decoherence-compatibility" I mean that the outcome of any collapse is some normalized branch vector, and the probability of this outcome occurring conforms to the Born Rule. A form of Ehrenfest's Theorem derived for open quantum systems implies that stochastic, localized trajectories defined through this effective, decoherence-compatible collapse prescription are overwhelmingly likely to be approximately Newtonian on certain timescales.  

\subsection{Setup} \label{Setup}

Decoherence-based models of classical behavior take the system of interest $S$  - say, the center of mass of a baseball - to be embedded in a larger closed system $SE$, where $E$ consists of all degrees of freedom external to $S$ and is typically referred to as $S$'s ``environment" (e.g., air molecules, photons, dust particles, measuring apparati, human observers). The Hilbert space $\mathcal{H}_{SE}$ of $SE$ is the tensor product of the Hilbert space $\mathcal{H}_{S}$ of $S$ and the Hilbert space $\mathcal{H}_{E}$ of $E$.  The state of $SE$, $| \Psi \rangle \in \mathcal{H}_{SE}$, is assumed always to evolve according to a Schrodinger equation of the form, 

\begin{align}
 i \frac{\partial | \Psi \rangle}{\partial t} = \left( \hat{H}_{S} \otimes \hat{I}_{E} + \hat{I}_{S} \otimes \hat{H}_{E} + \hat{H}_{I}  \right) | \Psi \rangle,
\end{align}

\noindent where we have set $\hbar \equiv 1$, $\hat{H}_{S}$ and $\hat{H}_{E}$ operate, respectively, on $\mathcal{H}_{S}$ and $\mathcal{H}_{E}$, $\hat{I}_{E}$ is the identity on $\mathcal{H}_{E}$ , $\hat{I}_{S}$ is the identity on $\mathcal{H}_{S}$, $\hat{H}_{I}$ operates on states in $\mathcal{H}_{S} \otimes \mathcal{H}_{E}$, and

\begin{equation}
\hat{H}_{S} = \frac{\hat{P}^{2}}{2M} + V(\hat{X}),
\end{equation}

\noindent with $\hat{X}$ and $\hat{P}$ the position and momentum operators on $\mathcal{H}_{S}$. A well-known model of this form is the model of quantum Brownian motion, in which $E$ consists of many independent oscillators whose position operators are linearly coupled through $\hat{H}_{I}$ to the position operator of $S$ (see, for example, \cite{JoosZehBook}, \cite{schlosshauer2008decoherence} and \cite{caldeira1983path}).
For the purposes of this discussion, I will assume only that $\hat{H}_{E}$ and $\hat{H}_{I}$ are consistent with a Caldeira-Leggett master equation for the reduced state $\hat{\rho} \equiv Tr_{E}[| \Psi \rangle \langle \Psi |]$ of $S$:

\begin{equation} \label{DecoherenceMaster}
i \frac{d \hat{\rho}}{dt} = [ \hat{H}_{S}, \hat{\rho} ]  - i \Lambda \left[ \hat{X},\left[ \hat{X}  , \hat{\rho} \right] \right],
\end{equation}

\noindent where $\Lambda$ is a constant that depends on the detailed parameters of $\hat{H}_{E}$ and $\hat{H}_{I}$. The first term on the right-hand side of this equation generates the unitary evolution of $\hat{\rho}$ while the second term accounts for the effects of decoherence on $\hat{\rho}$. 
For simplicity of illustration, I have assumed here that dissipative effects of the environment can be ignored. However, the sort of analysis of classical behavior given here can be straighforwardly generalized to the case where a dissipative term is included in the master equation for $\hat{\rho}$, the main difference being that the dynamical equations of the corresponding classical model will contain a frictional term in this case. For derivations of the Caldeira-Leggett equation in the context of different models for the unitary evolution of $SE$, see \cite{schlosshauer2008decoherence}, \cite{JoosZehBook} and references therein.

\subsection{Environmental Decoherence and Coherent Pointer States} \label{CoherentStateBasis}


Assume for the moment that the total system $SE$  begins at $t=0$ in a product state $| \Psi_{0} \rangle = | \psi \rangle \otimes | \phi \rangle$, where $| \psi \rangle \in \mathcal{H}_{S}$ and $| \phi\rangle \in \mathcal{H}_{E}$. Though the assumption of an initial product state is unrealistic in cases where the evolution of $SE$ is always unitary and where $S$ and $E$ interact, analyzing this simple case first is an important step in describing the evolution of more general initial states. In general, such a state will evolve under $SE$'s unitary dynamics into a state that is entangled. The rate at which the degree of entanglement between the two systems increases depends strongly on the choice of $| \psi \rangle$. Two popular ways to quantify the degree of entanglement between a system and its environment are the von Neumann entropy, $S_{VN} \equiv - Tr (\hat{\rho}  \log( \hat{\rho}))$ and the linear entropy, $S_{L} = 1 - Tr( \hat{\rho}^{2})$ (the latter is the leading approximation to the former when the former is expanded in powers of $(1 - \hat{\rho})$). Several authors have argued that for the systems of interest to us here, these measures of entanglement increase most slowly when $| \psi \rangle$ is a minimum uncertainty coherent state $| Q,P \rangle \in \mathcal{H}_{S}$ that is narrowly localized about some values of position $Q$ and momentum $P$. Expressed in a position basis for $S$, these states are Gaussian in form:

\begin{equation}
\Psi_{Q,P}(X) \equiv \langle X | Q, P \rangle = A e^{-\alpha(X-Q)^{2}} e^{iPX},
\end{equation}

\noindent where $A$ is an appropriate normalization constant, $\alpha$ determines the position- and momentum-space distribution widths associated with this state and these widths multiply to $\frac{\hbar}{2}$. Zurek, Habib and Paz have argued this point for the specific case where the potential $V$ in $S$'s self Hamiltonian is that of a harmonic oscillator.
\footnote{In fact, while we limit our attention here to the case of ``pure decoherence" described in (\ref{DecoherenceMaster}), these authors consider the somewhat more general case in which dissipation also plays a role in the system's dynamics and in which the oscillator is underdamped.
}
Through a method known as the \textit{predictability sieve}, they consider the time dependence of the linear entropy $S(t)$ for a range of different initial pure states $| \psi \rangle$ of $S$ and find that the $| \psi \rangle$ for which $S$ increases most slowly are minimum uncertainty coherent states \cite{zurek1993coherent}. More qualitative, heuristic arguments have been used to extend this conclusion to more general choices of potential $V$; see, for example, \cite{schlosshauer2008decoherence}, Ch.'s 2.8 and 5.2 and \cite{wallace2012emergent}, Ch. 3. Partly because minimum uncertainty coherent states are narrowly localized both in position and momentum and because they tend to avoid entanglement on longer timescales than other states, they are widely thought to provide a quantum mechanical counterpart to points in classical phase space.

Assuming that the pointer states of $S$ are coherent states, let us consider how this assumption constrains the unitary evolution of the quantum state of $SE$. My discussion here closely follows \cite{wallace2012emergent}, Section 3.7. Using the condensed notation $Z \equiv(Q,P)$, $ | Z \rangle \equiv | Q,P \rangle$, an arbitrary state of $SE$ at some initial time $t=0$ can be expressed in the form,

\begin{equation} \label{ArbInit}
| \chi_{0}  \rangle = \int dZ_{0} \ \alpha(Z_{0}) \ | Z_{0}  \rangle \otimes  | \phi(Z_{0}) \rangle
\end{equation}

\noindent where $\alpha(Z_{0})$ is an expansion coefficent, $| Z_{0}  \rangle \in \mathcal{H}_{S}$ and $ | \phi(Z_{0}) \rangle \in \mathcal{H}_{E}$.  (Note that, initially, the $| \phi(Z_{0}) \rangle$ need not be orthogonal.) 
After the very brief timescale $\tau_{D}$  associated with decoherence, we will have 
that $\langle \phi(Z'_{0}) | \phi(Z_{0}) \rangle \approx 0$ for $Z_{0}$ and $Z'_{0}$ sufficiently different. 
To determine the evolution of the overall state, let us first consider the unitary evolution of a single element of the initial superposition, $| Z_{0} \rangle \otimes | \phi(Z_{0}) \rangle$, and then use the linearity of the Schrodinger evolution to determine the evolution of the full state. On the timescale $\Delta t >> \tau_{D}$ over which $\hat{H}_{S}$ induces significant changes on $\mathcal{H}_{S}$, a single element of the above superposition evolves as 

\begin{equation}
  | Z_{0} \rangle \otimes | \phi(Z_{0}) \rangle  \overset{\Delta t}{\Longrightarrow}  \int dZ_{1}  \beta(Z_{0}, Z_{1}) | Z_{1} \rangle \otimes | \phi(Z_{0}, Z_{1}) \rangle
\end{equation}

\noindent where $\langle \phi(Z'_{0}, Z'_{1}) | \phi(Z_{0}, Z_{1}) \rangle \approx 0$ if $Z_{0}$ and $Z'_{0}$, or $Z_{1}$ and $Z'_{1}$,  are sufficiently different. 
Likewise, evolving each element of this last superposition forward in time by $\Delta t$, we have

\begin{equation}
  | Z_{1} \rangle \otimes | \phi(Z_{0}, Z_{1}) \rangle  \overset{\Delta t}{\Longrightarrow}  \int dZ_{2}  \beta(Z_{0},Z_{1},Z_{2}) | Z_{2} \rangle \otimes | \phi(Z_{0}, Z_{1},Z_{2}) \rangle,
\end{equation}

\noindent where $\langle \phi(Z'_{0}, Z'_{1}, Z'_{2}) | \phi(Z_{0}, Z_{1},Z_{2}) \rangle \approx 0$ if $Z_{k}$ and $Z'_{k}$ are sufficiently different, for any $0 \leq k \leq 2$. Iterating this process $N$ times, we obtain,

\begin{equation}
 | Z_{0} \rangle \otimes | \phi(Z_{0}) \rangle  \overset{N \Delta t}{\Longrightarrow}  \int dZ_{1}  ... dZ_{N} \ B(Z_{0}, ...,Z_{N}) \ | Z_{N} \rangle \otimes | \phi(Z_{0}, ..., Z_{N}) \rangle
 \end{equation}

\noindent where

\begin{equation} \label{DefinitionB}
B(Z_{0}, ...,Z_{N}) \equiv \beta(Z_{0}, Z_{1}) \ \beta(Z_{0}, Z_{1}, Z_{2}) \ ... \ \beta(Z_{0}, Z_{1},Z_{2},..., Z_{N-1},Z_{N}).
\end{equation}

\noindent By linearity of the Schrodinger evolution, we then have for the evolution of the arbitrary initial state (\ref{ArbInit}),

\begin{equation} \label{StateEvolution}
| \chi(N \Delta t) \rangle =    \int dZ_{0} ... dZ_{N} \ \alpha(Z_{0}) B(Z_{0},...,Z_{N}) \  | Z_{N} \rangle \otimes | \phi(Z_{0},...,Z_{N}) \rangle,
\end{equation}

\noindent with 

\begin{equation} \label{EnvStatesDec}
\langle \phi(Z'_{0},...,Z'_{N}) | \phi(Z_{0}, ...,Z_{N}) \rangle \approx 0 \ \text{ if } \ Z_{k} \ \text{and} \ Z'_{k} \ \text{ are sufficiently different, for any} \ 0 \leq k \leq N.
\end{equation}




\noindent Relations (\ref{StateEvolution}) and (\ref{EnvStatesDec}) provide a completely general expression for the evolution of the quantum state of $SE$ up to an arbitrary time $N \Delta t$ 
\footnote{Here $N \Delta t$ is implicitly assumed to be less than timescales associated with quantum Poincare recurrence and recoherence.
}
under the assumption that the coherent states $| Z \rangle$ are the pointer states of $S$ under its interaction with $E$.  Because the different environmental states $| \phi(Z_{0}, ...,Z_{N}) \rangle$ are mutually orthogonal for different trajectories $(Z_{0},...,Z_{N})$, each such environmental state serves in a sense to record the sequence or history of localized states of $S$ associated with this trajectory, $(| Z_{0} \rangle, | Z_{1} \rangle,...,| Z_{N} \rangle)$. Note that this trajectory, while localized, need not be approximately Newtonian.

\subsection{Environmental Decoherence and Decoherent Histories} \label{DecEnvHist}

The formalism of decoherent histories is useful for describing the branching structure that the quantum state acquires as a result of environmental decoherence. I briefly and informally review the basic elements of this framework, including the mathematical concepts of projection-valued measure (PVM), positive operator-valued measure (POVM), history operators, and decoherence of histories.



\subsubsection{PVM's and POVM's}

A projection-valued measure (PVM) on a Hilbert space $\mathcal{H}$ is a set of self-adjoint operators $\{ \hat{P}_{\alpha} \}$ on $\mathcal{H}$ such that

\begin{align}
& \sum_{\alpha} \hat{P}_{\alpha} = \hat{I}, \\
& \hat{P}_{\alpha} \hat{P}_{\beta} = \delta_{\alpha \beta} \hat{P}_{\alpha}, \label{PVMOrthog}
\end{align}

\noindent  where there is no  summation over  repeated indices. The concept of a positive operator-valued measure (POVM) on $\mathcal{H}$ generalizes the notion of a PVM by relaxing the requirement of orthogonality in (\ref{PVMOrthog}). Thus, a positive-operator-valued measure (POVM) on a Hilbert space $\mathcal{H}$ is a set $\{ \hat{\Pi}_{\alpha} \}$ of positive operators such that  
 
\begin{align}
& \sum_{\alpha} \hat{\Pi}_{\alpha} = \hat{I}.
\end{align}

\noindent An operator $\hat{O}$ is positive if it is self-adjoint and $\langle \Psi | \hat{O}|\Psi \rangle \geq 0 $ for every $|\Psi \rangle \in \mathcal{H}$. Note that every PVM is also a POVM.
\footnote{See \cite{nielsen2010quantum} or \cite{de2002foundations} for a more comprehensive introduction to PVM- and POVM-based treatments of quantum measurement.
}

\subsubsection{An Approximate PVM Constructed from Coherent States}

Following Ch. 3 of Wallace's \cite{wallace2012emergent}, consider a partition $\{ \mu_{\alpha} \}$ of the classical phase space $\Gamma_{S}$ of the system $S$ introduced in Section \ref{Setup}. To such a partition we can associate the POVM $\{ \hat{\Pi}_{\alpha} \}$ on $\mathcal{H}_{S}$, where the operators $\hat{\Pi}_{\alpha}$ are defined by

\begin{equation} \label{POVMDefinition}
\hat{\Pi}_{\alpha} \equiv \int_{\mu_{\alpha}} dZ \ |Z \rangle \langle Z|.
\end{equation}

\noindent Assuming that the cells $\mu_{\alpha} = u_{\alpha^{x}} \times u_{\alpha^{p}}$ of the partition have configuration space volume $d_{X} \equiv vol(u_{\alpha^{x}})$ and momentum space volume $d_{P} \equiv vol(u_{\alpha^{p}}) $ larger than those associated with a coherent pointer state, 
the operators in this POVM will also constitute an approximate PVM, so that

\begin{equation} \label{POVMOrthog}
\hat{\Pi}_{\alpha} \hat{\Pi}_{\beta} \approx \delta_{\alpha \beta} \hat{\Pi}_{\alpha}.
\end{equation}

\noindent We can then extend the set $\{ \hat{\Pi}_{\alpha} \}$ to an approximate PVM on $\mathcal{H}_{SE}$ by defining the set of operators $\{ \hat{P}_{\alpha} \}$, where $\hat{P}_{\alpha} \equiv \hat{\Pi}_{\alpha} \otimes \hat{I}_{E}$, with $\hat{I}_{E}$ the identity on $\mathcal{H}_{E}$. It is worth emphasizing that the classical phase space $\Gamma_{S}$ employed in defining this approximate PVM is not assumed to have any fundamental ontological status, but rather is employed simply as a mathematical tool in defining a structure that is fundamentally quantum mechanical. The fact that this particular PVM is useful for analyzing the branching structure of the pure state evolution of $SE$ follows from the fact that the coherent states are pointer states of $S$ under its interaction with $E$, which in turn follows from the form of $SE$'s quantum Hamiltonian.

\subsubsection{Decoherent Histories} \label{DecoherentHistories}

The approximate coherent state PVM just defined is useful for describing the branching structure that emerges from the unitary quantum state evolution through decoherence. Let us consider the evolution of $SE$'s state $| \Psi \rangle$ up to some arbitrary time $t >> \tau_{D}$, dividing $t$ into $N$ equal intervals $\Delta t >> \tau_{D}$, so that  $t=N \Delta t$:

\begin{align} \label{Histories}
| \Psi(N \Delta t)\rangle &=  e^{- i\hat{H} N \Delta t} | \Psi_{0} \rangle \\
& =  \left( \sum_{\alpha_{N}} \hat{P}_{\alpha_{N}} \right) e^{- i \hat{H} \Delta t} \left(\sum_{\alpha_{N-1}} \hat{P}_{\alpha_{N-1}} \right) ...  e^{-i \hat{H} \Delta t} \left(\sum_{\alpha_{1}} \hat{P}_{\alpha_{1}} \right) e^{- i \hat{H} \Delta t} \left(\sum_{\alpha_{0}} \hat{P}_{\alpha_{0}} \right)|  \Psi_{0}  \rangle \\
& = \sum_{\alpha_{0},...,\alpha_{N}}    \hat{P}_{\alpha_{N}} e^{-\frac{i}{\hbar} \hat{H} \Delta t}   \hat{P}_{\alpha_{N-1}}    ...   e^{-i \hat{H} \Delta t} \hat{P}_{\alpha_{1}} e^{-\frac{i}{\hbar}\hat{H} \Delta t} \hat{P}_{\alpha_{0}} |  \Psi_{0}  \rangle     \\
& =  \sum_{\alpha_{0},...,\alpha_{N}} \hat{C}_{\alpha_{0}...\alpha_{N}} |  \Psi_{0}  \rangle 
\end{align} 

\noindent where 
$ \hat{C}_{\alpha_{0}...\alpha_{N}} \equiv \hat{P}_{\alpha_{N}}  e^{-\frac{i}{\hbar}\hat{H} \Delta t}\hat{P}_{\alpha_{N-1}} ...  e^{-i \hat{H} \Delta t} \hat{P}_{\alpha_{1}} e^{-\frac{i}{\hbar}\hat{H} \Delta t} \hat{P}_{\alpha_{0}} $ and in going from the first to the second line we have used the fact that $\sum_{\alpha_{i}} \hat{P}_{\alpha_{i}} = \hat{I}_{SE}$. Each component  $\hat{C}_{\alpha_{0} ... \alpha_{N}} |  \Psi_{0}  \rangle$ corresponds to a particular history or sequence $(\mu_{\alpha_{0}},...,\mu_{\alpha_{N}})$ of regions through phase space. I will further assume that the same partition $\{ \mu_{\alpha} \}$ is used to define the set of approximate PVM operators \{ $\hat{P}_{\alpha_{i}}$ \} at all times $i \Delta t$, where $0 \leq i \leq N$.  The two histories $(\alpha_{0},...,\alpha_{N})$ and $(\alpha'_{0},...,\alpha'_{N})$ are said to be \textit{decoherent} if the corresponding Hilbert space vectors are orthogonal - that is, if

\begin{equation} \label{HistoriesDecoherence}
\langle \Psi_{0} | \hat{C}_{\alpha'_{0} \alpha'_{1} ... \alpha'_{N}}^{ \dagger} \hat{C}_{\alpha_{0} \alpha_{1}...\alpha_{N}}   | \Psi_{0}\rangle \approx 0 
\end{equation}

\noindent for $\alpha_{k} \neq \alpha'_{k}$ for any $0 \leq k \leq N$ (this condition is sometimes designated ``medium decoherence" in the decoherent histories literature). The set of all histories $\{ (\alpha_{0},...,\alpha_{N}) \}$ is called decoherent if any two distinct histories in the set are decoherent. A history is said to be \textit{realized} if its corresponding Hilbert space vector has non-negligible weight - that is, if $ | \hat{C}_{\alpha_{0} ... \alpha_{N}} |  \Psi_{0}  \rangle | \geq \epsilon$ for some small threshold value $\epsilon$. Any history that is not realized is automatically decoherent with respect to every other history, since the inner product of the associated component of the quantum state vector with any other such component will be zero. As Paz and Zurek have argued, and as one can see for oneself by expanding the vectors $\hat{C}_{\alpha_{0}...\alpha_{N}} |  \Psi_{0}  \rangle$ in terms of coherent states using the definition (\ref{POVMDefinition}) of the operators $\hat{\Pi}_{\alpha_{i}}$,  environmental decoherence relative to a coherent state pointer basis for $S$, as expressed in (\ref{EnvStatesDec}), entails decoherence of different phase space histories, as expressed in (\ref{HistoriesDecoherence}) \cite{paz1993environment}.

\subsection{Decoherence, Branching and the Unitary Quantum State Evolution} \label{UnitaryBranching}

If the condition (\ref{HistoriesDecoherence}) is satisfied at all times $N \Delta t$ for different $N$, the unitary evolution considered at successive time steps takes the form, 

\begin{flalign*}
& | \Psi(0) \rangle = | \Psi_{0} \rangle & 
\end{flalign*}

\begin{flalign*}
& | \Psi(\tau_{D}) \rangle = \sum_{\alpha_{0} } \hat{C}_{\alpha_{0}} | \Psi_{0} \rangle & \\
&& \langle \Psi_{0}|\hat{C}^{\dagger}_{\alpha'_{0}}   \hat{C}_{\alpha_{0} } | \Psi_{0} \rangle \approx 0 \ if \ \alpha_{0} \neq \alpha'_{0}
\end{flalign*}
\begin{flalign*}
& | \Psi(\Delta t) \rangle = \sum_{\alpha_{0}, \alpha_{1}} \hat{C}_{\alpha_{0} \alpha_{1}} | \Psi_{0} \rangle & \\
&& \langle \Psi_{0}|\hat{C}^{\dagger}_{\alpha'_{0} \alpha'_{1}}   \hat{C}_{\alpha_{0} \alpha_{1}} | \Psi_{0} \rangle \approx 0 \ if \ \alpha_{i} \neq \alpha'_{i}  \ for \ 0 \leq i \leq 1 
\end{flalign*}
\begin{flalign*}
& | \Psi(2\Delta t) \rangle = \sum_{\alpha_{0}, \alpha_{1}, \alpha_{2}} \hat{C}_{\alpha_{0} \alpha_{1} \alpha_{2}} | \Psi_{0} \rangle   \\
&& \langle \Psi_{0}|\hat{C}^{\dagger}_{\alpha'_{0} \alpha'_{1} \alpha'_{2}}   \hat{C}_{\alpha_{0} \alpha_{1} \alpha_{2}} | \Psi_{0} \rangle \approx 0 \ if \ \alpha_{i} \neq \alpha'_{i}  \ for \ 0 \leq i \leq 2 
\end{flalign*}
\ \ \ \ \ \ \ \ \ \  \vdots 
\begin{flalign*}
 | \Psi(N \Delta t) \rangle = \sum_{\alpha_{0},\alpha_{1}, \alpha_{2}, ... , \alpha_{N}} &\hat{C}_{\alpha_{0} \alpha_{1} \alpha_{2}... \alpha_{N}} | \Psi_{0} \rangle  &  \ \ \ \ \ \ \ \ \ \ \ &  \ \ \ \ \ \ & \ \ \ \ \ \  & \ \ \ \  \\
& \  \langle \Psi_{0}|\hat{C}^{\dagger}_{\alpha'_{0} \alpha'_{1} \alpha'_{2}...\alpha'_{N}}   \hat{C}_{\alpha_{0} \alpha_{1} \alpha_{2}...\alpha_{N}} | \Psi_{0} \rangle \approx 0 \  if \ \alpha_{i} \neq \alpha'_{i} \ for \ 0 \leq i \leq N, \\
\end{flalign*}

\noindent where $\hat{C}_{\alpha_{0}} \equiv e^{- i \hat{H} \tau_{D}} \hat{P}_{\alpha_{0}}$.
\footnote{Note that it is necessary to evolve the total state forward by the decoherence time $\tau_{D}$ before the decoherence condition can be expected to hold among components associated with different $\alpha_{0}$.
}
The unitary evolution in this case exhibits a tree-like structure in that the probability that one of the mutually decohered components $\hat{C}_{\alpha_{0}  ... \alpha_{i} } | \Psi_{0} \rangle $ of the total quantum state at time $i \Delta t$ transitions into the component $\hat{C}_{\beta_{0} ... \beta_{i} \beta_{i+1}} | \Psi_{0} \rangle$ at time $(i+1)\Delta t$ is effectively zero unless $(\beta_{0}, ... , \beta_{i}) = (\alpha_{0}, ... , \alpha_{i})$
\footnote{To calculate the transition amplitude, apply the evolution operator $e^{-i \hat{H} \Delta t }$ to the vector $\hat{C}_{\alpha_{0} ... \alpha_{i}} | \Psi_{0} \rangle$, followed by the identity operator in the form $\sum_{\alpha_{i+1}} \hat{P}_{\alpha_{i+1}}$, yielding $ \sum_{\alpha_{i+1}} \hat{C}_{\alpha_{0} ... \alpha_{i} \alpha_{i+1}} | \Psi_{0} \rangle$. Take the inner product of the resultant vector with $\hat{C}_{\beta_{0} ... \beta_{i} \beta_{i+1}} | \Psi_{0} \rangle$ for arbitrary $(\beta_{0}, ... , \beta_{i}, \beta_{i+1})$. Decoherence of histories at time $(i+1)\Delta t$ ensures that this inner product will be zero unless $(\beta_{0}, ... , \beta_{i}) = (\alpha_{0}, ... , \alpha_{i})$.
}
. That is, the unitary evolution of the vector $\hat{C}_{\alpha_{0} ... \alpha_{i}} | \Psi_{0} \rangle$ will give rise to multiple distinct, mutually decohered vectors $\hat{C}_{\alpha_{0} ... \alpha_{i} \alpha_{i+1}} | \Psi_{0} \rangle$ (one for each $\alpha_{i+1}$) at future times, but the initial segments of the histories associated with those future vectors must coincide with the history $(\alpha_{0}, ... ,\alpha_{i})$ of their parent vector.
\footnote{For a more detailed discussion of branching, see  \cite{griffiths1993consistent} and \cite{wallace2012emergent}, Ch. 3.
}
For this reason, the $\hat{C}_{\alpha_{0} ... \alpha_{i}} | \Psi_{0} \rangle$ are sometimes called \textit{branch vectors} and the corresponding normalized states \textit{branch states}. 
Let us further stipulate that these branch vectors should be maximally fine-grained with respect to $S$ in the sense that the phase space regions $\mu_{\alpha}$ used to define the approximate coherent state PVM $\{\hat{P}_{\alpha} \}$ should be as small as possible consistent with (\ref{POVMOrthog}) and with different histories $(\alpha_{0},  ...  , \alpha_{i})$ being mutually decohered at each time $i \Delta t$. In this case, the dimensions $d_{X}$ and $d_{P}$ of the partition cells $\mu_{\alpha}$ should not be much greater, respectively, than the position- and momentum-space widths of a coherent pointer state, and the phase space volume $d_{X} d_{P}$ should not be much larger than $1$ - or, equivalently, $\hbar^{n}$ (where, recall, we have set $\hbar \equiv 1$). Note that  $\hbar^{n} = 1$ represents the volume in $S$'s phase space associated with a single coherent state, assuming that $S$'s phase space has dimension $2n$.



\subsection{Effective, Stochastic, Quasiclassical State Evolution} \label{SectionFAPPCollapse}

Within the decoherence literature, the world of experience is widely thought to be described at any given time by just one of the quasi-classical branches that arise in the unitary evolution of $SE$'s total quantum state. To each branch at each time, we can formally assign an effective, normalized ``branch state",

\begin{equation}
\frac{1}{W_{\alpha_{0}...\alpha_{N}}}\hat{C}_{\alpha_{0} ... \alpha_{N}} | \Psi_{0} \rangle,
\end{equation}

\noindent where $W_{\alpha_{0} ... \alpha_{N}} \equiv  \sqrt{ \langle \Psi_{0}|\hat{C}^{\dagger}_{\alpha_{0} ...\alpha_{N}}   \hat{C}_{\alpha_{0} ... \alpha_{N}} | \Psi_{0} \rangle} $. Given the unitary branching evolution just described, we may formally define the following effective, stochastic evolution for the branch state of $SE$:

\begin{flalign} \label{BranchSequence}
& \frac{1}{W_{\alpha_{0}}} \hat{C}_{\alpha_{0}} | \Psi_{0} \rangle \xrightarrow{prob. \frac{|W_{\alpha_{0} \alpha_{1}}|^{2}}{|W_{\alpha_{0}}|^{2}}}    \frac{1}{W_{\alpha_{0} \alpha_{1}} } \hat{C}_{\alpha_{0} \alpha_{1}} | \Psi_{0} \rangle \xrightarrow{prob. \frac{|W_{\alpha_{0} \alpha_{1} \alpha_{2}}|^{2}}{|W_{\alpha_{0} \alpha_{1}}|^{2}}}    \frac{1}{W_{\alpha_{0} \alpha_{1} \alpha_{2}} } \hat{C}_{\alpha_{0} \alpha_{1} \alpha_{2}} | \Psi_{0} \rangle  \xrightarrow{prob. \frac{|W_{\alpha_{0} \alpha_{1} \alpha_{2} \alpha_{3}}|^{2}}{|W_{\alpha_{0} \alpha_{1} \alpha_{2}}|^{2}}} \nonumber \\ 
& \ \nonumber \\
&\ \ ... \ \ \xrightarrow{prob. \frac{|W_{\alpha_{0} \alpha_{1} \alpha_{2} ... \alpha_{N-1} \alpha_{N}}|^{2}}{|W_{\alpha_{0} \alpha_{1}  \alpha_{2}... \alpha_{N-1}}|^{2}}}  \frac{1}{W_{\alpha_{0} \alpha_{1} \alpha_{2} ...\alpha_{N}}}  \hat{C}_{\alpha_{0} \alpha_{1} ... \alpha_{N}} | \Psi_{0} \rangle \ \ \xrightarrow{prob. \frac{|W_{\alpha_{0} \alpha_{1} \alpha_{2} ... \alpha_{N-1} \alpha_{N} \alpha_{N+1}}|^{2}}{|W_{\alpha_{0} \alpha_{1}  \alpha_{2}... \alpha_{N-1} \alpha_{N}}|^{2}}} ...  .
\end{flalign} 

\noindent Each step in this evolution, 

\begin{flalign} \label{FAPPCollapse2}
& \frac{1}{W_{\alpha_{0}...\alpha_{i}}}  \hat{C}_{\alpha_{0} ... \alpha_{i}} | \Psi_{0} \rangle \xrightarrow{prob. \ \frac{|W_{\alpha_{0}...\alpha_{i} \alpha_{i+1}}|^{2}}{|W_{\alpha_{1}...\alpha_{i}}|^{2}}} \frac{1}{W_{\alpha_{1}...\alpha_{i} \alpha_{i+1} } } \hat{C}_{\alpha_{1} ... \alpha_{i} \alpha_{i+1}} | \Psi_{0} \rangle .
\end{flalign}

\noindent may be regarded as a combination of a unitary, deterministic branching process,
  
\begin{flalign} \label{FAPPSupp}
& \frac{1}{W_{\alpha_{0}...\alpha_{i}}}  \hat{C}_{\alpha_{0} ... \alpha_{i}} | \Psi_{0} \rangle \xrightarrow{e^{-i\hat{H}\Delta t}}  \frac{1}{W_{\alpha_{0}...\alpha_{i}}} \sum_{\alpha_{i+1}} \hat{C}_{\alpha_{0} ... \alpha_{i} \alpha_{i+1}} | \Psi_{0} \rangle,
\end{flalign}

\noindent where $\langle \Psi_{0}|\hat{C}^{\dagger}_{\alpha'_{0} ...\alpha'_{i} \alpha'_{i+1} }   \hat{C}_{\alpha_{1} ...\alpha_{i} \alpha_{i+1}} | \Psi_{0} \rangle \approx 0$ if $\alpha_{j} \neq \alpha'_{j}$ for any $0 \leq k \leq i+1$, and a non-unitary, stochastic  collapse process, 

\begin{flalign}  \label{FAPPCollapse1}
&\frac{1}{W_{\alpha_{0}...\alpha_{i}}} \sum_{\alpha_{i+1}} \hat{C}_{\alpha_{0} ... \alpha_{i} \alpha_{i+1}} | \Psi_{0} \rangle
  \xrightarrow{prob. \frac{|W_{\alpha_{0}...\alpha_{i} \alpha_{i+1}}|^{2}}{|W_{\alpha_{0}...\alpha_{i}}|^{2}}}  \frac{1}{W_{\alpha_{0}...\alpha_{i} \alpha_{i+1} } }  \hat{C}_{\alpha_{0} ... \alpha_{i} \alpha_{i+1}} | \Psi_{0} \rangle. 
\end{flalign}

\noindent It is straightforward to see that the collapse probability $ \frac{|W_{\alpha_{0}...\alpha_{i} \alpha_{i+1}}|^{2}}{|W_{\alpha_{0}...\alpha_{i}}|^{2}}$ coincides with the Born Rule. 
Though \textit{ad hoc}, this collapse prescription is more precise than the prescription appearing in most quantum mechanics texts in that it is explicitly required to be decoherence-compatible - that is, the only allowed outcomes of a collapse process under this prescription are branch states defined by decoherence, whereas the more conventional textbook prescription makes no mention of decoherence
\footnote{See, for example, \cite{merzbacherQM}, \cite{sakurai1995modern} and \cite{griffiths2005introduction}. Schwabl's \cite{schwabl2007quantum} is something of an exception in that it includes a short discussion of decoherence-based models of quantum measurement (thanks to Michel Janssen for pointing me to Schwabl's text).
}. 
It is also important to note that the square modulus $|W_{\alpha_{1}, \alpha_{2}, ... , \alpha_{i}}|^{2}$ of the branch state $\hat{C}_{\alpha_{1} \alpha_{2} ... \alpha_{i}} | \Psi_{0} \rangle$ is equal to the probability of the sequence  $\frac{1}{W_{\alpha_{0}}}\hat{C}_{\alpha_{0}} | \Psi_{0} \rangle$, $\frac{1}{W_{\alpha_{0} \alpha_{1}}} \hat{C}_{\alpha_{0} \alpha_{1}} | \Psi_{0} \rangle$, ... , $\frac{1}{W_{\alpha_{0} \alpha_{1} ... \alpha_{i-1}}}\hat{C}_{\alpha_{0} \alpha_{1} ... \alpha_{i-1}} | \Psi_{0} \rangle$, $\frac{1}{W_{\alpha_{0} \alpha_{1} ... \alpha_{i-1} \alpha_{i}}}\hat{C}_{\alpha_{0} \alpha_{1} ... \alpha_{i-1} \alpha_{i}} | \Psi_{0} \rangle$ on our effective stochastic evolution
\footnote{To see this, recall that the square modulus of $\hat{C}_{\alpha_{1} \alpha_{2} ... \alpha_{N}} | \Psi_{0} \rangle$ is equal, by definition, to the quantity $|W_{\alpha_{1} \alpha_{2} ... \alpha_{N}}|^{2}$. Let us now independently calculate the probability of the sequence of states in (\ref{BranchSequence}) according to our effective branch-relative state evolution. This probability is simply equal to the product of the probabilities $\frac{|W_{\alpha_{1}, \alpha_{2}, ... , \alpha_{i}, \alpha_{i+1}}|^{2}}{|W_{\alpha_{1}, \alpha_{2}, ... , \alpha_{i}}|^{2}}$ for all transitions $(\alpha_{1}, \alpha_{2}, ... , \alpha_{i}) \rightarrow (\alpha_{1}, \alpha_{2}, ... , \alpha_{i}, \alpha_{i+1})$: 

\begin{align}
&\frac{|W_{\alpha_{1} \alpha_{2}}|^{2}}{|W_{\alpha_{1}}|^{2}} \times \frac{|W_{\alpha_{1} \alpha_{2} \alpha_{3}}|^{2}}{|W_{\alpha_{1} \alpha_{2}}|^{2}} \times ... \times \frac{|W_{\alpha_{1} \alpha_{2} ... \alpha_{N-2} \alpha_{N-1}}|^{2}}{|W_{\alpha_{1} \alpha_{2} ... \alpha_{N-2}}|^{2}} \times \frac{|W_{\alpha_{1} \alpha_{2} ... \alpha_{N-1} \alpha_{N}}|^{2}}{|W_{\alpha_{1} \alpha_{2} ... \alpha_{N-1}}|^{2}} \\
& \ \\
& = |W_{\alpha_{1} \alpha_{2} ... \alpha_{N}}|^{2}.
\end{align} 

\noindent By a series of simple cancellations, we can see that the probability on the left-hand side is equal to the square modulus of the branch weight  $|W_{\alpha_{1} \alpha_{2} ... \alpha_{N}}|^{2}$. 
}.
Thus, the unitary branching structure described in Section \ref{UnitaryBranching} formally encodes every possible 
\footnote{At the coarse level of description where interpretation-specific details of collapse are omitted, there is a correspondingly coarse sense of ``possibility" on which future descendents of one's current branch all represent distinct possible future evolutions, and on which branches other than the one that happens to have been realized could have possibly been realized instead. Of course, the precise sense in which other branches could have been realized, and in which future descendents of one's current branch represent possible future evolutions, varies across interpretations. In dBB theory, for example, the future branch is completely predetermined by an exact specification of the Bohmian configuration and the quantum state; thus, the sense in which other branches are possible requires us to imagine alternative values for the present Bohmian configurations. In GRW theory, by contrast, there is a sense in which the future branch-relative evolution is completely open and not determined by any details of the present state of the world. It is the coarse, effective sense of possibility, rather than the more precise sense associated with any particular interpretation, that I invoke in the present context. 
}
evolution allowed by our effective, stochastic branch-state evolution as well as the probability of every such evolution.

The stochastic evolution (\ref{BranchSequence}) implies a corresponding evolution of the branch-relative reduced density matrix of $S$. Defining
${\scriptstyle \hat{\rho}_{\alpha_{0}...\alpha_{n}} \equiv \frac{1}{|W_{ \alpha_{0}...\alpha_{n}} |^{2}} Tr_{E}\left( \hat{C}_{\alpha_{0} ... \alpha_{n}} | \Psi_{0} \rangle \langle \Psi_{0} | \hat{C}^{\dagger}_{\alpha_{0} ... \alpha_{n}} \right)}$, this evolution is 

\begin{flalign} \label{DensityOperatorBranchSequence}
& \hat{\rho}_{\alpha_{0}} \xrightarrow{}   \hat{\rho}_{\alpha_{0} \alpha_{1}} \xrightarrow{}    \hat{\rho}_{\alpha_{0} \alpha_{1} \alpha_{2}}  \xrightarrow{} \ ...  \ \xrightarrow{}   \hat{\rho}_{\alpha_{0} \alpha_{1} \alpha_{2} ... \alpha_{N}} \xrightarrow{} \ ... .
\end{flalign}


\noindent The states $\hat{\rho}_{\alpha_{0}...\alpha_{i}}$ are quasi-classical in that the distribution $\rho_{\alpha_{1}...\alpha_{i}}(X) \equiv  \langle X | \hat{\rho}_{\alpha_{1}...\alpha_{i}}| X\rangle$ over $S$'s position is narrowly peaked on the scale $d_{X}$ while the distribution  $\tilde{\rho}_{\alpha_{1}...\alpha_{i}}(P) \equiv \langle P | \hat{\rho}_{\alpha_{1}...\alpha_{i}}| P \rangle $ over $S$'s momentum is narrowly peaked on the scale $d_{P}$ (recall that $d_{X}$ and $d_{P}$ are the position- and momentum-space dimensions of the partition cell $\mu_{\alpha_{i}}$). 
To each such state, there corresponds a unique past sequence $\hat{\rho}_{\alpha_{0}}, \hat{\rho}_{\alpha_{0} \alpha_{1}}, ... ,  \hat{\rho}_{\alpha_{0} \alpha_{1} ... \alpha_{i-1}}  $ of quasi-classical states from which it must have evolved according to the effective stochastic dynamics, but many possible future sequences of branch-relative quasi-classical states to which it may give rise. Note that the discreteness of the timesteps $\Delta t$  
is an artefact of the manner in which we have chosen to describe an evolution that occurs continuously in time. Defining the stochastic trajectory $\hat{\rho}_{\alpha}(N \Delta t)  \equiv  \hat{\rho}_{\alpha_{0} ... \alpha_{N}}$, where $\hat{\rho}_{\alpha_{0} ... \alpha_{N}}$ is the outcome of the $N^{th}$ ``collapse", we can see that when we make the time intervals $\Delta t$ successively smaller - and $N$ successively larger for fixed $t$ - our discrete, stochastic, reduced-state trajectory provides successively better approximations to a continuous, stochastic quasiclassical trajectory of reduced
\footnote{Note that ``reduced" here is intended in the sense of ``reduced density matrix" - that is, to connote a tracing over environmental degrees of freedom.
}
states, $\hat{\rho}_{\alpha}(t)$. 

\subsection{Quantum Trajectories in Classical Phase Space} \label{BRPhaseSpace}

The sequence (\ref{DensityOperatorBranchSequence})  induces a stochastic trajectory on the classical phase space of $S$ when we take branch-relative expectation values of $S$'s position and momentum operators: 

\begin{align}
 X_{\alpha_{0}...\alpha_{N}} & \equiv \frac{1}{|W_{\alpha_{0}...\alpha_{N}}|^{2}} \langle \Psi_{0} |    \hat{C}^{\dagger}_{\alpha_{0}...\alpha_{N}} \left( \hat{X} \otimes \hat{I}_{E} \right)  \hat{C}_{\alpha_{0}...\alpha_{N}} | \Psi_{0} \rangle = Tr_{S} \left(  \hat{\rho}_{\alpha_{0}...\alpha_{N}} \hat{X} \right) \nonumber  \\  
  & \ \ \nonumber \\ 
 P_{\alpha_{0}...\alpha_{N}} & \equiv \frac{1}{|W_{\alpha_{0}...\alpha_{N}}|^{2}} \langle \Psi_{0} |    \hat{C}^{\dagger}_{\alpha_{0}...\alpha_{N}} \left( \hat{P} \otimes \hat{I}_{E} \right)  \hat{C}_{\alpha_{0}...\alpha_{N}} | \Psi_{0} \rangle = Tr_{S} \left( \hat{\rho}_{\alpha_{0}...\alpha_{N}} \hat{P}   \right), \nonumber 
\end{align} 

\noindent or, more concisely, 

\begin{empheq}[box=\widefbox]{align}
 Z_{\alpha_{0}...\alpha_{N}} & \equiv \frac{1}{|W_{\alpha_{0}...\alpha_{N}}|^{2}} \langle \Psi_{0} |    \hat{C}^{\dagger}_{\alpha_{0}...\alpha_{N}} \left( \hat{Z} \otimes \hat{I}_{E} \right)  \hat{C}_{\alpha_{0}...\alpha_{N}} | \Psi_{0} \rangle = Tr_{S} \left( \hat{\rho}_{\alpha_{0}...\alpha_{N}} \hat{Z}  \right)  \nonumber \\ 
\end{empheq}

\noindent where $ Z_{\alpha_{0}...\alpha_{N}} \equiv ( X_{\alpha_{0}...\alpha_{N}},  P_{\alpha_{0}...\alpha_{N}}) \in \mu_{\alpha_{N}}$ and $\hat{Z} \equiv (\hat{X} , \hat{P})$. The trajectory in classical phase space induced by the stochastic evolution (\ref{DensityOperatorBranchSequence}) is  

\begin{flalign*}
& Z_{\alpha_{0}} \ \   \xrightarrow{}   \ \ Z_{\alpha_{0} \alpha_{1}}  \ \   \xrightarrow{}   \ \ Z_{\alpha_{0} \alpha_{1} \alpha_{2}} \ \ \xrightarrow{}  \ \ ... \ \ \xrightarrow{}  \ \ Z_{\alpha_{0} \alpha_{1} \alpha_{2}...\alpha_{N}} \ \ \xrightarrow{} \ \ ... \ \ . 
\end{flalign*}


\noindent As in Section \ref{SectionFAPPCollapse}, we may take the time intervals $\Delta t$ successively smaller in order to approximate a continuous trajectory $Z^{q}(t) \equiv Tr_{S}[\hat{\rho}_{\alpha}(t) \hat{Z}]$, where $Z^{q}(N \Delta t) = Z_{\alpha_{0} ... \alpha_{N}} = Tr_{S}[\hat{\rho}_{\alpha_{0} ... \alpha_{N}} \hat{Z}]$.

\subsection{Why \textit{Newtonian} Trajectories? Dynamical Requirements for Inter-Model Reduction} \label{NewtonianTrajectories}


Within the decoherence literature, it is often thought that the determinate, quasi-classical states of affairs that characterize the world of our experience are represented by individual branches of the total quantum state. Thus, it is natural to suppose that the physical degrees of freedom that are represented in classical models by a point in phase space should be represented in the corresponding quantum model by branch-relative expectation values of position and momentum, around which the branch-relative distributions in these quantities are tightly peaked. Assuming that this is the case, there remains the further question of why, in cases where the system $S$ is the center of mass of a body such as the moon, a baseball, or an alpha particle, the evolution of these branch-relative expectation values according to the effective, stochastic quantum evolution described above should, with very high likelihood, approximate some solution to Newton's equations of motion. 

We can pose the question more formally and precisely within the context of the local, empirical, model-based approach to reduction described in Section \ref{InterModelReduction}. 
The state space of the high-level model $M_{h}$ in this case is the classical phase space $\Gamma_{S}$ of the system $S$. The high-level dynamics are prescribed by the Hamiltonian/Newtonian equations of motion for some choice of potential $V$. 
We can take the state space of the low-level model to be the space $\mathcal{Q}(\mathcal{H}_{S})$ of density matrices on the Hilbert space of the system $S$ and the low-level dynamics to be given by (\ref{DensityOperatorBranchSequence}), which in turn is derived from the prescription (\ref{BranchSequence}). 
\footnote{While the distinction between a physical system and its representations by different models plays an important role in the discussion here, I employ a slight abuse of terminology and use S to refer to both the physical system itself and its representations in quantum and classical models. It should be clear from context which usage is intended. 
}
Let $d_{c}$ denote the domain of states in $\Gamma_{S}$ such that Hamiltonian trajectories with initial conditions in $d_{c}$ track $S$'s behavior within margin of error  $\delta_{Z}$ over a timescale greater than $\tau_{c}$. As noted in Section \ref{InterModelReduction}, there will be some flexibility in the values $\delta_{Z}$, $\tau_{c}$ and $d_{c}$, which parametrize the fit between the classical model and the behavior of the system $S$. 
Fixing values for $(\delta_{Z}, \tau_{c}, d_{c})$, let $Z^{c}(t, Z_{0})$ represent the deterministic classical trajectory with initial condition $Z_{0} \in d_{c}$ and $Z^{q}(t, \hat{\rho}_{0})$ represent a phase space trajectory induced by the stochastic quantum evolution of the initial state $\hat{\rho}_{0}$. Furthermore, assume that $ Z_{0} =  Tr_{S} \left[ \hat{\rho}_{0} \hat{Z}\right]$, so that both the classical and induced quantum trajectories have the same starting point in $\Gamma_{S}$ at $t=0$. Following the discussion of Section \ref{InterModelReduction}, $reduction_{M}$ of our deterministic classical model to our effective stochastic quantum model requires that for each $Z_{0} \in d_{c}$, there exist a $\hat{\rho}_{0} \in \mathcal{Q}(\mathcal{H}_{S})$ such that $Tr_{S}[\hat{\rho}_{0} \hat{Z}] = Z_{0}$ and, with probability $1- \epsilon$ for very small $\epsilon$, 


\begin{empheq}[box=\widefbox]{align} \label{QCApproxZ}
\bigg| Z^{q}( t, \hat{\rho}_{0}) - Z^{c}( t, Z_{0})  \bigg| < 2 \delta_{Z} 
\end{empheq}

\noindent or, more explicitly,
 
\begin{empheq}[box=\widefbox]{align} \label{QCApproxZExplicit}
\left| Tr_{S} \left[ \hat{\rho}_{\alpha}(t) \hat{Z}\right]  - Z^{c}\left(t,Z_{0} \right) \right| < 2 \delta_{Z} 
\end{empheq}

\noindent for $ 0 \leq t < \tau_{c}$. Denote by $d_{q}$the set of all such $\hat{\rho}_{0}$. In this example, the bridge function $B$ appearing in Eq. (\ref{ComRed1}) is given by $B(\hat{\rho}) =  Tr_{S} \left[  \hat{\rho} \hat{Z} \right]$. 
We will see below that the domain $d_{q} \subset  \mathcal{Q}(\mathcal{H}_{S})$ consists of states $\hat{\rho}_{0}$ that are narrowly peaked in both their position and momentum ensemble widths. 
We now set ourselves to outlining an argument as to why condition (\ref{QCApproxZ}) should hold under circumstances where classical behavior is known to occur, so that the quantum model gives at least as accurate (if not necessarily as convenient) a representation of the system's behavior as the classical model in these cases.


\subsection{Ehrenfest's Theorem for Open Quantum Systems} \label{EhrenfestOpen}

Our approach to showing that condition (\ref{QCApproxZ}) holds in relevant cases lies in a certain extension of Ehrenfest's Theorem to open quantum systems that was first derived (as far as I know) by Joos and Zeh (see \cite{JoosZehBook}, Ch.3). In order to highlight the novel aspects of this approach, let us begin by reviewing the well-known attempt to recover classical behavior based on the much more commonly cited closed-system form of Ehrenfest's Theorem. 

In its familiar form, Ehrenfest's Theorem applies only to cases where the system of interest $S$ is closed and in a pure state, and where its evolution is always governed by a Schrodinger equation with Hamiltonian $\hat{H} = \frac{\hat{P}^{2}}{2M} + V(\hat{X})$. The theorem, which follows immediately from taking the expectation value of the Heisenberg equation of motion for momentum, states that given these assumptions, the following relation holds for all times $t$ and all states $| \psi \rangle$:

\begin{equation}
\frac{d \langle  \hat{P} \rangle}{dt} =  - \langle   \hat{\frac{\partial V(X) }{ \partial X}}\rangle. 
\end{equation}

\noindent Combined with the relation $\frac{d \langle \hat{X} \rangle}{dt} =   \langle  \frac{\hat{P}}{M}  \rangle$ (the expectation of the Heisenberg equation for position), this implies 
$\frac{d^{2} \langle  \hat{X}  \rangle}{dt^{2}} =  - \langle  \hat{\frac{\partial V(X) }{ \partial X}}  \rangle$. Note that despite its formal resemblance to Newton's second law of motion, this last relation does \textit{not} entail that expectation values of position and momentum evolve approximately classically. For that to be the case, it is necessary that $\frac{d^{2} \langle  \hat{X}  \rangle}{dt^{2}} =  - \frac{\partial V( \langle \hat{X} \rangle) }{ \partial \langle \hat{X} \rangle}= \frac{\partial V(X) }{ \partial X}\bigg|_{\langle \hat{X} \rangle}$.  If we impose the restriction that $ | \psi \rangle$ be a wave packet whose spatial width is narrow by comparison with the characteristic length scales of the potential $V$, it follows that

\begin{equation} \label{EhrenfestNewton}
\frac{d \langle \hat{P} \rangle}{dt} \approx  - \frac{\partial V(X) }{ \partial X} \bigg|_{\langle \hat{X}  \rangle},
\end{equation}

\noindent which, together with the relation $\frac{d \langle  \hat{X} \rangle }{dt} = \frac{\langle  \hat{P} \rangle}{M}$, entails that this condition is satisfied approximately.  The relation (\ref{EhrenfestNewton}), in turn, can typically only be expected to hold for as long as $|\psi(X)|^{2}$ remains suitably narrowly peaked by comparison with the dimensions of $V$. Except in the special case where $V$ is a harmonic oscillator potential - in which case the relation (\ref{EhrenfestNewton}) is satisfied with exact equality for all states at all times, and coherent states maintain their width - quantum mechanical wave packets can generally be expected to spread out over time. 

For systems that are open, as is the case for real systems that exhibit classical behavior, the assumption that $S$ is always in a pure state and that the dynamics of $S$'s state are always unitary does not apply. Taking quantum mechanics as a universal theory, interaction of such systems with microscopic degrees of freedom in their environments is unavoidable, so that our quantum model of $S$ must allow for the possibility of entanglement with its environment $E$. Taking the state of $S$ then to be given by the reduced density matrix $\hat{\rho}$ and to evolve according to the canonical master equation (\ref{DecoherenceMaster}), and using the relations $\langle \hat{P} \rangle = Tr [{\hat{\rho} \hat{P}}] $ and $\langle \hat{X} \rangle = Tr [{\hat{\rho} \hat{X}}] $, Joos and Zeh have shown that

\begin{equation}
\frac{d \ Tr_{S}[\hat{\rho}  \hat{P}] }{dt} =  -  Tr_{S}\left[ \hat{\rho} \hat{\frac{\partial V(X) }{ \partial X}}\right],
\end{equation}

\noindent or more concisely,

\begin{equation} \label{OpenEhrenfest}
\frac{d \langle \hat{P} \rangle}{dt} =  - \langle \hat{\frac{\partial V(X) }{ \partial X}} \rangle,
\end{equation}

\noindent (see \cite{JoosZehBook}, Ch.3). If we now impose the restriction to density matrices $\hat{\rho}$ for which the position-space ensemble width - that is, the width of the distribution $ \langle X | \hat{\rho}| X \rangle$ - is small by comparison with the characteristic length scales of $V$, it follows that, as in the case where $S$ is closed,

\begin{equation} \label{OpenEhrenfestApprox}
\frac{d \langle \hat{P} \rangle}{dt} \approx  - \frac{\partial V(X) }{ \partial X}\bigg|_{\langle \hat{X} \rangle}.
\end{equation}

\noindent Combined with the relation $\frac{d \langle \hat{X} \rangle }{dt} = \frac{\langle  \hat{P} \rangle}{M}$, this entails that the total expectation values (as opposed to branch-relative expectation values)  $\langle \hat{X} \rangle$ and $\langle \hat{P} \rangle$ will follow an approximately Newtonian trajectory as long as the distribution $\rho(X) =   \langle X | \hat{\rho}| X \rangle$ remains suitably narrow by comparison with the dimensions of $V$. Thus, the timescales on which the total expectation values $\langle \hat{X} \rangle$ and $\langle \hat{P} \rangle$ follow an approximately Newtonian trajectory will depend on the rate at which the distribution $\rho(X)$ spreads over time.

\subsection{Implications of the Open-Systems Ehrenfest's Theorem for the Branching Structure of the Quantum State} \label{EhrenfestImplications}

Let us now consider how the open-systems version of Ehrenfest's Theorem constrains the branching structure of the total quantum state of $SE$ and, in turn, the stochastic quasi-classical trajectories for $S$ derived from this branching structure through the decoherence-compatible collapse prescription described in Section \ref{SectionFAPPCollapse}. As we will see, the theorem implies that these trajectories will, with very high likelihood, be approximately Newtonian over the timescale where ensemble spreading
\footnote{Recall that the ensemble width of the reduced density matrix of $S$ in some basis $\{ | A \rangle\}$ (usually position or momentum) is the width of the distribution  $\rho_{A}(A) =   \langle A | \hat{\rho}| A \rangle$. Thus, it represents the spread along the diagonal of this reduced density matrix. By contrast, the coherence width of the same reduced density matrix in this basis represents the spread of the density matrix perpendicular to the diagonal and quantifies the range of coherence in the basis $\{ | A \rangle\}$. Decoherence with respect to a given basis typically leads to a rapid reduction of the coherence width of the reduced density matrix in that basis, assuming that off-diagonal elements are not already miniscule.  
}
of $S$'s reduced state remains below a certain threshhold; reduction of the classical to the quantum model requires that this timescale be greater than or equal to the timescale over which Newtonian trajectories successfully track the behavior of the physical system $S$. 

To begin, let us relate the total (as opposed to branch-relative) evolution of $S$'s reduced density matrix $\hat{\rho}(t)$ - which is constrained by the open-systems Ehrenfest's Theorem - to the unitary branching structure of the total pure state $| \Psi(t) \rangle$ of $SE$. The relation $\hat{\rho}(N \Delta t) = Tr_{E} \left(| \Psi(N \Delta t) \rangle \langle  \Psi(N \Delta t) | \right)$, together with the substitution $| \Psi(N \Delta t) \rangle = \sum_{\alpha_{0}, ... , \alpha_{N}} \hat{C}_{\alpha_{0} ... \alpha_{N}} | \Psi_{0} \rangle$ and the assumption of environmental decoherence relative to a coherent state pointer basis for $S$ yields

{\small
\begin{align}\label{DensityBranching}
\hat{\rho}(N \Delta t) & =  \sum_{\bar{\alpha}, \bar{\beta}}  Tr_{E} \left( \hat{C}_{\alpha_{0} ... \alpha_{N}} | \Psi_{0} \rangle  \langle \Psi_{0} | \hat{C}^{\dagger}_{\beta_{0} ... \beta_{N}}  \right) \\
& \approx \sum_{\bar{\alpha}} Tr_{E} \left( \hat{C}_{\alpha_{0} ... \alpha_{N}} | \Psi_{0} \rangle  \langle \Psi_{0} | \hat{C}^{\dagger}_{\alpha_{0} ... \alpha_{N}}  \right) \\
& = \sum_{\bar{\alpha}}  |W_{\alpha_{0} ... \alpha_{N}}|^{2} \ \hat{\rho}_{\alpha_{0} ... \alpha_{N}}, 
\end{align}
}

\noindent where $\hat{\rho}_{\alpha_{0} ... \alpha_{N}} \equiv \frac{1}{|W_{\alpha_{0} ... \alpha_{N}}|^{2} } Tr_{E} \left( \hat{C}_{\alpha_{0} ... \alpha_{N}} | \Psi_{0} \rangle  \langle \Psi_{0} | \hat{C}^{\dagger}_{\alpha_{0} ... \alpha_{N}}  \right)$, $\bar{\alpha} \equiv (\alpha_{0}, ... , \alpha_{N})$ and $\bar{\beta} \equiv (\beta_{0}, ... , \beta_{N})$. The second equality above follows from the relation  $Tr_{E} \left( \hat{C}_{\bar{\alpha}} | \Psi_{0} \rangle  \langle \Psi_{0} | \hat{C}^{\dagger}_{\bar{\beta}}  \right) \approx 0$ for $\bar{\alpha} \neq \bar{\beta}$, which in turn is a consequence of environmental decoherence. Note that it is specifically environmental decoherence, and not simply decoherence of histories, that is needed to enforce this result. 

\footnote{We can see that $Tr_{E} \left( \hat{C}_{\bar{\alpha}} | \Psi_{0} \rangle  \langle \Psi_{0} | \hat{C}^{\dagger}_{\bar{\beta}}  \right) \approx 0$ for $\bar{\alpha} \neq \bar{\beta}$ by expanding the vectors $\hat{C}_{\alpha_{1} ... \alpha_{N}} | \Psi_{0} \rangle$ in terms of coherent states using the definition of the operators $\hat{\Pi}_{\alpha_{i}}$ in (\ref{POVMDefinition}):
\begin{align}
\hat{C}_{\alpha_{1} ... \alpha_{N}} | \Psi_{0} \rangle & \equiv \int_{\mu_{\alpha_{0}}}  ... \int_{\mu_{\alpha_{N}}} dz_{0} ...  dz_{N} \  \hat{C}_{z_{0} ... z_{N}} | \chi_{0} \rangle \\
&= \int_{\mu_{\alpha_{0}}}  ... \int_{\mu_{\alpha_{N}}} dz_{0}  ...  dz_{N}  \ | z_{N} \rangle \otimes | \tilde{\xi}(z_{0},...,z_{N}) \rangle \\
&=  \int_{\mu_{\alpha_{0}}}  ... \int_{\mu_{\alpha_{N}}} dz_{0}  ...  dz_{N} \  w(z_{0},...,z_{N}) \ | z_{N} \rangle \otimes | \phi(z_{0},...,z_{N}) \rangle
\end{align}
\noindent where $\hat{C}_{z_{0} ... z_{N}}  \equiv  \left(  | z_{N} \rangle  \langle z_{N}  | \otimes \hat{I}_{E}  \right) e^{-i\hat{H} \Delta t} ...   \left(  | z_{1} \rangle  \langle z_{1}  | \otimes \hat{I}_{E}  \right)  e^{-i\hat{H} \Delta t} \left(  | z_{0} \rangle  \langle z_{0}  | \otimes \hat{I}_{E}  \right) =  | z_{N} \rangle \otimes | \tilde{\xi}(z_{0}, ... ,z_{N}) \rangle$, $ | \tilde{\xi}( z_{0}, ..., z_{N}) \rangle \equiv  \sum_{i} | e_{i} \rangle  \langle z_{N}, e_{i}  | e^{-i\hat{H}\Delta t} \hat{C}_{z_{0}  ... z_{N-1}} \ | z_{0}\rangle = w(z_{0},...,z_{N}) \ | \phi(z_{0},...,z_{N}) \rangle$, $\{ | e_{i} \rangle \}$ is any basis of $\mathcal{H}_{E}$,  $w(z_{0},...,z_{N}) \equiv  \sqrt{\langle \tilde{\xi}(z_{0},...,z_{N})  | \tilde{\xi}(z_{0},...,z_{N}) \rangle}$ and $ | \phi(z_{0},...,z_{N}) \rangle \equiv \frac{ | \tilde{\xi}( z_{0}, ..., z_{N}) \rangle}{w(z_{0},...,z_{N})}$   (the tilde over  $\xi$ denotes the fact that this vector is not normalized). From the assumption of environmental decoherence expressed in Eq. (\ref{EnvStatesDec}), we see that the environmental vectors $| \tilde{\xi}(z_{0},...,z_{N}) \rangle$ associated with different histories $\bar{\alpha}$ and $\bar{\beta}$ must be mutually orthogonal. From this it follows that
\begin{align}
Tr_{E} \left( \hat{C}_{\bar{\alpha}} | \Psi_{0} \rangle  \langle \Psi_{0} | \hat{C}^{\dagger}_{\bar{\beta}}  \right) & = \sum_{i} \langle e_{i} | \hat{C}_{\bar{\alpha}} | \Psi_{0} \rangle  \langle \Psi_{0} | \hat{C}^{\dagger}_{\bar{\beta}} | e_{i} \rangle \\
&= \int_{\bar{\mu}^{N}_{\alpha}} dz^{N} \int_{\bar{\mu}^{N}_{\beta}}  dz'^{N} \left( \sum_{i}\langle e_{i} |  \tilde{\xi}(z_{0},...,z_{N}) \rangle \langle \tilde{\xi}(z'_{0},...,z'_{N}) | e_{i} \rangle \right) | z_{N} \rangle \langle z'_{N} | \\
&= \int_{\bar{\mu}^{N}_{\alpha}} dz^{N} \int_{\bar{\mu}^{N}_{\beta}}  dz'^{N}  \left(    \langle \tilde{\xi}(z'_{0},...,z'_{N})    |  \tilde{\xi}(z_{0},...,z_{N}) \rangle   \right) | z_{N} \rangle \langle z'_{N} | \\
& \approx 0 
\end{align}
\noindent where $\int_{\bar{\mu}^{N}_{\alpha}} dz^{N} \equiv \int_{\mu_{\alpha_{1}}} dz_{1} ... \int_{\mu_{\alpha_{N}}} dz_{N} $ and likewise for $\int_{\bar{\mu}^{N}_{\beta}} dz'^{N}$. The last equality comes from the orthogonality of environment vectors $|  \tilde{\xi}(z_{0},...,z_{N}) \rangle$ associated with distinct histories $\bar{\alpha}$ and $\bar{\beta}$.  
}

Letting $\rho(X, t) \equiv  \langle X | \hat{\rho}(t)| X \rangle$ and $\rho_{\alpha_{0} ... \alpha_{N}}(X) \equiv \langle X | \hat{\rho}_{\alpha_{0} ... \alpha_{N}}  | X \rangle$, this last relation  expressed in the position representation is

\begin{equation} \label{PositionDensityMatrix}
\rho(X, N \Delta t) = \sum_{\alpha_{0}, ... , \alpha_{N}}  |W_{\alpha_{0} ... \alpha_{N}}|^{2} \ \rho_{\alpha_{0} ... \alpha_{N}}(X).
\end{equation}

\noindent Each component $\rho_{\alpha_{0} ... \alpha_{N}}(X)$ is narrowly peaked within the spatial volume $d_{X}$ associated with the phase space region $\mu_{\alpha_{N}}$ and is localized around  $\langle \hat{X} \rangle \equiv Tr_{S}[\hat{\rho}_{\alpha_{0} ... \alpha_{N}} \hat{X}]$, which lies in the narrow set of configurations $\mu_{\alpha^{x}_{N}}$ associated with  $\mu_{\alpha_{N}}$. 
From (\ref{PositionDensityMatrix}), then, we can see that the spatial ensemble distribution $\rho(X,t)$ of $S$ at each time is a superposition of non-interfering, potentially overlapping narrow wave packets $\rho_{\alpha_{0} ... \alpha_{N}}(X)$, each associated with a branch of the overall quantum state and contributing to $\rho(X,N \Delta t)$  in proportion to the squared weight $|W_{\alpha_{0} ... \alpha_{N}}|^{2}$ of this branch. Over a single timestep $\Delta t$, a single branch packet may evolve into a superposition of different branch packets, so that $\rho_{\alpha_{0} ... \alpha_{N}}(X) \rightarrow \sum_{\alpha_{i}}\frac{ |W_{\alpha_{0} ... \alpha_{N} \alpha_{N+1}}|^{2}}{|W_{\alpha_{0} ... \alpha_{N}}|^{2}}  \rho_{\alpha_{0} ... \alpha_{N} \alpha_{N+1}}(X) $.  To each branch $(\alpha_{0}, ...  , \alpha_{N})$, one can associate a unique past trajectory for the localized wave packet of $S$: $\rho_{\alpha_{0}}(X) \rightarrow \rho_{\alpha_{0} \alpha_{1}}(X) \rightarrow ... \rightarrow \rho_{\alpha_{0} \alpha_{1} ... \alpha_{N}}(X) $.  Under the stochastic, effective evolution associated with the branching structure of the quantum state, this evolution has a probability $|W_{\alpha_{0} ... \alpha_{N}}|^{2}$ of occurring, and the state of the system $S$ at an arbitary time $N \Delta t$ is always one of the quasi-classical branch packets $ \rho_{\alpha_{0} ... \alpha_{N}}(X)$. 

Taking the evolution $\rho(X,t)$ as given, we can use the relation (\ref{PositionDensityMatrix}) to determine which histories $(\alpha_{0},...,\alpha_{N})$ are realized in the unitary evolution of the quantum state $| \Psi \rangle \in \mathcal{H}_{SE}$ and which are not.  Since the functions $\rho_{\alpha_{0} ... \alpha_{N}}(X)$ are all positive-valued, we can see that $|W_{\alpha_{0} ... \alpha_{N}}|^{2} \approx 0$ if $\rho_{\alpha_{0} ... \alpha_{N}}(X)$ is localized outside of the support
\footnote{Here, I use the term ``support" to designate the subset of a function's domain over which the function's value is non-negligible - so, greater in magnitude than some appropriately small $\epsilon$ - rather than the subset of its domain where the function's value is non-zero, as required by the strict mathematical definition of support. 
}
of $\rho(X, N \Delta t)$
 - that is, if the spatial part $u_{\alpha_{N}^{x}}$ of $\mu_{\alpha_{N}}$ lies outside of the region where $\rho(X, N \Delta t)$ is non-negligible. What's more, $|W_{\alpha_{0} ... \alpha_{N}}|^{2} \approx 0$ for any history $(\alpha_{0}, ... ,\alpha_{N})$ such that $\rho_{\alpha_{0}...\alpha_{i}}(X)$ lies outside the support of $\rho(X, i \Delta t)$ for any $1 \leq i \leq N$. That is, the only histories that are realized in the unitary evolution $SE$'s pure state at a given time are those for which the associated sequence of past branch packets of $S$ lies in the support of $\rho(X,t)$ at all earlier times. 
 \footnote{This can be seen as follows. Assume that $\rho_{\alpha_{0}...\alpha_{i}}(X)$ lies outside of the support of $\rho(X, i \Delta t)$ for some $1 \leq i \leq N$. Then $|W_{\alpha_{0} ... \alpha_{i}}|^{2} \approx 0$. But $|W_{\alpha_{0} ... \alpha_{i} ... \alpha_{N}}|^{2} \leq |W_{\alpha_{0} ... \alpha_{i}}|^{2}$, since no branch vector can have weight greater than that of a branch vector from which it is descended. Therefore, $|W_{\alpha_{0} ... \alpha_{i} ... \alpha_{N}}|^{2} \approx 0$ for all $N \geq i$. 
 }
We will see that combined with this reasoning, the open-systems version of Ehrenfest's Theorem implies that on timescales where the spatial ensemble width of an initially narrow wave packet remains below a certain margin, all branch-relative trajectories with non-negligible branch weight are approximately Newtonian. On the stochastic, quasi-classical evolution associated with this branching structure, it is therefore overwhelmingly likely that the quantum mechanical phase space trajectory of the system in question will be approximately Newtonian.


To see this in detail, let us begin by assuming that  $SE$'s initial pure state $| \Psi_{0} \rangle$ is a product state of the form $| \Psi_{0} \rangle =  | Z_{0} \rangle \otimes | \phi_{0} \rangle$, so that  $\hat{\rho}(0) = | Z_{0} \rangle \langle Z_{0} |$ and the initial ensemble distribution $\rho(X, 0) = |\langle X| Z_{0} \rangle |^{2}$ is a narrow wave packet localized on the scale of a coherent pointer state. We will later use the linearity of the Schrodinger evolution and the fact that the states $| Z_{0} \rangle$ constitute a basis for $\mathcal{H}_{S}$ to extend our analysis to more general choices of $| \Psi_{0} \rangle$.
By virtue of the reduced dynamics of $S$, the initially narrow distribution $\rho(X, 0)$ will spread out over time.  
Denoting by $H_{Z_{0}}^{N}$ the set of histories $(\alpha_{0},...,\alpha_{N})$ up to time $N \Delta t$ such that the associated branch packets $\rho_{\alpha_{0}...\alpha_{i}}(X)$ are localized inside the support of $\rho(X, i \Delta t)$ for all $0 \leq i \leq N$,  we may approximate (\ref{PositionDensityMatrix}) by restricting the sum to histories in $H_{Z_{0}}^{N}$, since $|W_{\alpha_{0} ... \alpha_{N}}|^{2} \approx 0$ for all other histories: 

\begin{equation} \label{BranchPacketsRestricted}
\rho(X, N \Delta t) \approx \sum_{\alpha_{0}, ... , \alpha_{N} \in H_{Z_{0}}^{N}}  |W_{\alpha_{0} ... \alpha_{N}}|^{2} \ \rho_{\alpha_{0} ... \alpha_{N}}(X).
\end{equation}

\noindent 
Morever, because $|W_{\alpha_{0} ... \alpha_{N}}|^{2} \approx 0$ for any $(\alpha_{0},...,\alpha_{N}) \notin H_{Z_{0}}^{N}$, it follows that the pure state $| \Psi(N \Delta t) \rangle $ of $SE$ is approximated by,
 
\begin{equation} \label{BranchingApprox}
| \Psi(N \Delta t) \rangle \approx   \sum_{\alpha_{0}, ... , \alpha_{N} \in H_{Z_{0}}^{N}} \hat{C}_{\alpha_{0} ... \alpha_{N}} \left( | z_{0} \rangle \otimes | \phi(z_{0}) \rangle \right).  
\end{equation}

\noindent The effective, stochastic, quasiclassical phase space evolution associated with this branching structure will give rise with very high likelihood to trajectories in $H_{Z_{0}}^{N}$. However, the question remains as to why these trajectories should be approximately Newtonian.

To answer this question, recall that expectation values $\langle \hat{Z} \rangle \equiv \left(\langle \hat{X} \rangle,\langle \hat{P} \rangle \right) = \left( Tr_{S} \left[ \hat{\rho}(t)\hat{X}\right], Tr_{S} \left[ \hat{\rho}(t)\hat{P} \right] \right)$ with respect to the total reduced state $\hat{\rho}$ (as opposed to the branch-relative reduced state $\hat{\rho}_{\alpha}$) will evolve in approximately Newtonian fashion as long as the ensemble width $\Delta X(t)$ associated with the distribution $\rho(X,t)$ remains below the characteristic length scale $l_{V}$ on which the potential $V$ varies. 
Note also that the position and momentum ensemble widths $\Delta X(t)$ and  $\Delta P(t)$ - where the latter is the width of the distribution $\tilde{\rho}(P,t) \equiv \langle P | \hat{\rho}(t) | P \rangle$ -  quantify the average deviation of branch-relative trajectories away from the trajectory $\langle \hat{Z} \rangle(t)$. Therefore, trajectories in  $H_{Z_{0}}^{N}$ will approximate the Newtonian trajectory $Z^{c}(t,Z_{0})$ within margin $2 \delta_{Z}$, as the condition (\ref{QCApproxZ}) requires, if $\Delta Z(t) \equiv \left (\Delta X(t), \Delta P(t) \right) \leq 2 \delta_{Z} $ and $\Delta X(t) \leq l_{V}$ - or, equivalently, if $\Delta Z(t) \leq \left( \min(2 \delta_{X}, l_{V}), 2 \delta_{P} \right)$.
Let $T$ be the timescale up to which this condition holds. For $reduction_{M}$ to occur, it must be the case that $T \geq \tau_{c}$. That is, the timescale on which the trajectory $\langle \hat{Z} \rangle(t)$ remains approximately Newtonian and branch-relative trajectories are likely to remain tightly concentrated around this trajectory must be greater than or equal to the timescale over which the exact Newtonian trajectory tracks the behavior of the system. Otherwise, branch-relative quantum trajectories would not track the system's behavior in all cases where classical ones do, as $reduction_{M}$ requires. 

\begin{figure}[] \centering 
\includegraphics[scale=.30]{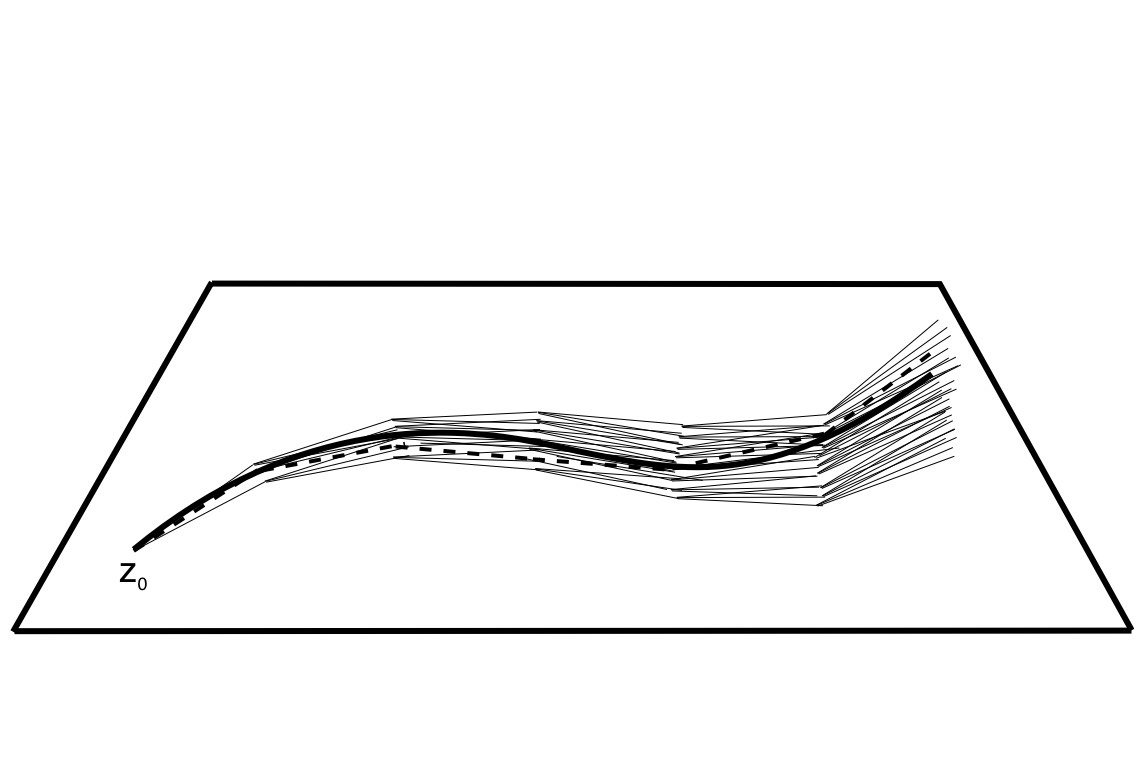} 
\caption{\footnotesize Assuming that $SE$ begins in a state of the form $| Z_{0}  \rangle \otimes  | \phi(Z_{0}) \rangle$, the ensemble widths $\Delta X$ and $\Delta P$ of $S$'s reduced state will spread. If $\Delta X$ remains suitably small relative to the characteristic scale of spatial variation of the potential $V$, expectation values of position and momentum will evolve along a trajectory that approximates the Newtonian trajectory associated with initial condition $Z_{0}$ (the exact Newtonian trajectory is represented by the thick black line). Since $\Delta X$ and $\Delta P$ also quantify the average deviation of branch-relative phase space trajectories from $\langle \hat{X} \rangle$ and $\langle \hat{P} \rangle$, branch-relative trajectories (thin black lines) will remain close to the Newtonian trajectory on timescales where these ensemble widths remain suitably small. 
The effective, stochastic, quasiclassical evolution associated with this branching structure singles out one of these branch-relative trajectories (dotted line) as the determinate evolution of the system.
}
\label{fig:NewtonBranching}
\end{figure}

Assuming for the moment that $T \geq \tau_{c}$, it follows that the set $H_{Z_{0}}^{N}$ in (\ref{BranchingApprox}) will consist only of histories $(\alpha_{0},...,\alpha_{N})$ that lie within $2 \delta_{Z}$ of $Z^{c}(t,Z_{0})$, for all $N$ such that $N \Delta t < \tau_{c}$. It follows immediately that the stochastic phase space trajectory $Z^{q}(t)$ associated with this branching structure will very likely be approximately Newtonian within margin $2 \delta_{Z}$ over timescale $\tau_{c}$.  Defining $\Delta Z_{\alpha_{0} ... \alpha_{i}} \equiv Z_{\alpha_{0} ... \alpha_{i}} - Z^{c}(i \Delta t, Z_{0})$, we may decompose the quantum phase space trajectory of $S$ into a deterministic classical part $Z^{c}(i \Delta t, Z_{0})$ and a smaller stochastic part $\Delta Z_{\alpha_{0} ... \alpha_{i}}$ consisting of branch-specific ``quantum fluctuations": $Z^{q}(i \Delta t, |Z_{0}  \rangle \langle Z_{0} |) = Z_{\alpha_{0} ... \alpha_{i}} =  Z_{c}(i \Delta t, Z_{0})+ \Delta Z_{\alpha_{0} ... \alpha_{i}}$. Formulated in this manner, the stochastic quantum phase space evolution takes the form, 

{\footnotesize
\begin{align} \label{PhaseSpaceDecomp}
& Z_{0} \longrightarrow{} Z_{c}(\Delta t, Z_{0})+ \Delta Z_{\alpha_{0} \alpha_{1}}   \longrightarrow{} Z_{c}( 2 \Delta t, Z_{0})+ \Delta Z_{\alpha_{0} \alpha_{1} \alpha_{2}}  \longrightarrow{}  ... \longrightarrow{}  Z_{c}( N \Delta t, Z_{0})+ \Delta Z_{\alpha_{0} \alpha_{1} \alpha_{2} ... \alpha_{N} }  \longrightarrow{} ...
\end{align}
}

\noindent where  

\begin{equation} \label{PhaseSpaceFluct}
| \Delta Z_{\alpha_{0}...\alpha_{i}}| < 2\delta_{Z}
\end{equation} 

\noindent with probability $1- \epsilon$ for very small $\epsilon$, for all $i$ such that $i \Delta t < \tau_{c}$  (see Figure ~\ref{fig:NewtonBranching}). The fluctuations $\Delta Z_{\alpha_{1} ...\alpha_{i}}$ can be ignored for all practical purposes when they are sufficiently small relative to the degree of precision with which one cares to describe the system.  

Thus far, we have explained how the open-systems form of Ehrenfest's Theorem leads to the recovery of approximately Newtonian trajectories under the assumption that $| \Psi_{0} \rangle =  | Z_{0} \rangle \otimes | \phi_{0} \rangle$. Because the coherent states $ | Z_{0} \rangle$ form a basis (more precisely, an overcomplete basis) for $\mathcal{H}_{S}$, and because of the linearity of the Schrodinger evolution of $SE$, it is straightforward to extend our discussion to more general $|\Psi_{0} \rangle$. An arbitrary initial pure state $|\Psi_{0} \rangle$ of $SE$ can be expressed as $|\Psi_{0} \rangle = \int dZ_{0} \alpha(Z_{0}) | Z_{0} \rangle \otimes | \phi(Z_{0}) \rangle $ for some  $\alpha(Z_{0})$ and  $| \phi(Z_{0}) \rangle$ (where, initially, the $| \phi(Z_{0}) \rangle$ need not be mutually orthogonal for different $Z_{0}$). 
From  (\ref{BranchPacketsRestricted}) and the linearity of the reduced state evolution of $S$, it follows that the reduced state of $S$ at time $N \Delta t$ takes the form,

\begin{equation} 
\rho(X, N \Delta t) \approx \int dZ_{0} \ |\alpha(Z_{0})|^{2} \sum_{\alpha_{0}, ... , \alpha_{N} \in H_{Z_{0}}^{N}}  |W_{\alpha_{0} ... \alpha_{N}}|^{2} \ \rho_{\alpha_{0} ... \alpha_{N}}(X).
\end{equation}

\noindent The form of $SE$'s pure state will likewise be

\begin{equation}
| \Psi(N \Delta t) \rangle \approx \int dZ_{0} \  \alpha(Z_{0}) \left[ \sum_{\alpha_{1}, ... , \alpha_{N} \in H_{Z_{0}}^{N}} \hat{C}_{\alpha_{1} ... \alpha_{N}} \left( | Z_{0} \rangle \otimes | \phi(Z_{0}) \rangle \right) \right]. 
\end{equation}

\begin{figure}[] 
\includegraphics[scale=.23]{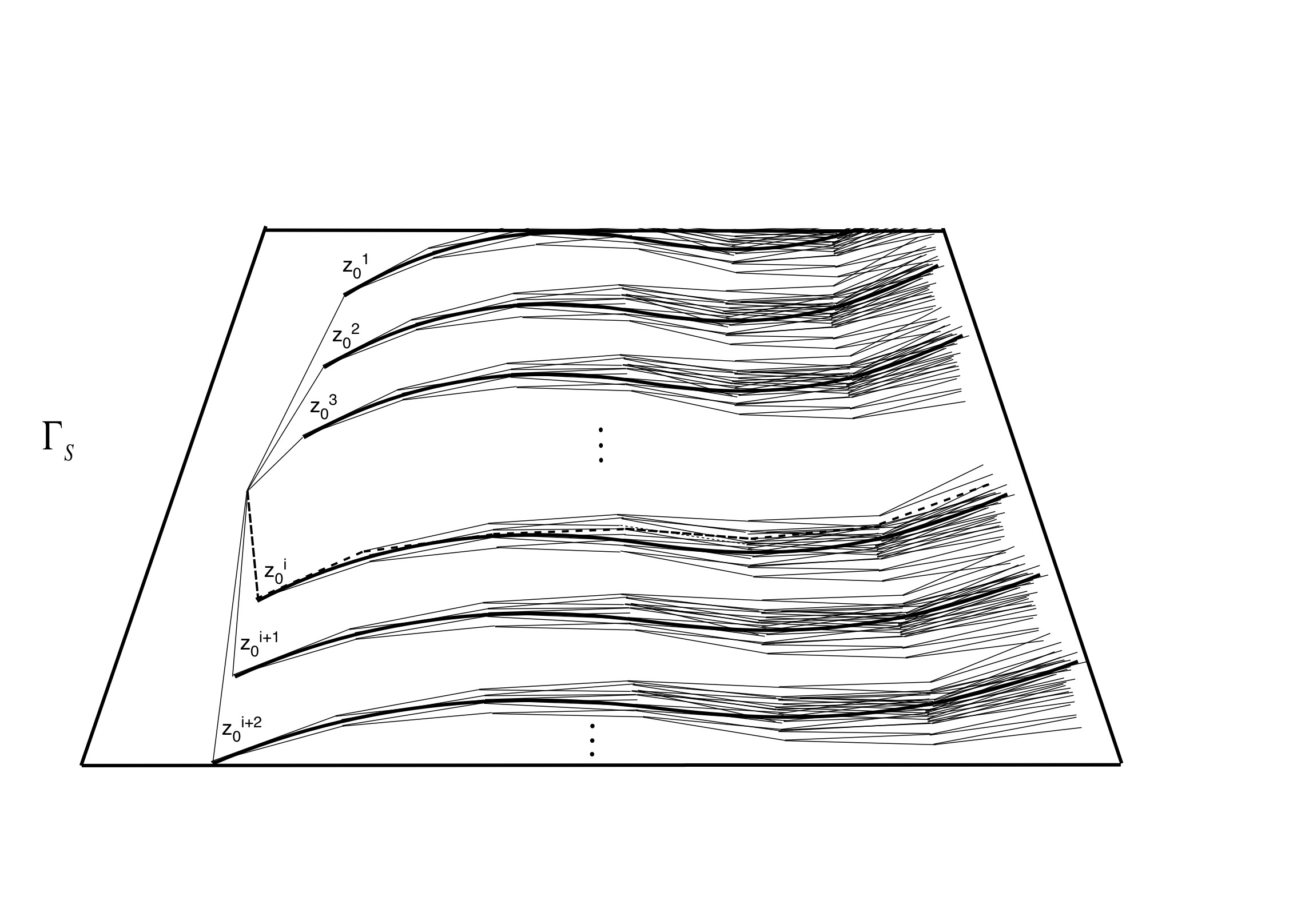}
\caption{\footnotesize A generic initial state for $SE$, which may be a coherent superposition of macroscopically different values for $S$'s position and momentum, will rapidly evolve under the unitary evolution into an incoherent superposition of mutually orthogonal, quasiclassical branch states associated with different initial conditions $Z_{0}$ in phase space. Each of these quasi-classical branch states gives rise to a set of branch-relative phase space trajectories (thin black lines) which approximate the Newtonian trajectory associated with the initial condition $Z_{0}$ (thick black lines) over the timescale $T$ defined in the main text. The stochastic quasiclassical evolution associated with this branching process acts on the initial coherent superposition by collapsing onto a single quasi-classical branch state associated with some particular value of $Z_{0}$ - in the diagram, $Z_{0}^{i}$. Thereafter, the stochastic phase space trajectory of $S$ (dotted line) fluctuates closely around the Newtonian trajectory associated with initial condition $Z_{0}^{i}$ until the ensemble width of position or momentum associated with the initial $Z_{0}^{i}$-branch state spreads beyond the required error threshold. }
 \label{fig:MultipleNewton}
\end{figure}

\noindent 
The effective, stochastic quasi-classical evolution associated with this branching structure can be described in qualitative terms as follows: the arbitary initial state $|\Psi_{0} \rangle$, which may contain coherent superpositions of macroscopically separated wave packets, collapses with probability $|\alpha(Z_{0})|^{2}$ onto a quasi-classical branch state relative to which $S$ is localized around $| Z_{0} \rangle$. Thereafter, it is overwhelmingly likely to evolve along a quasi-classical trajectory that closely approximates the Newtonian trajectory $Z^{c}(t, Z_{0})$, until time $T$ when the ensemble widths associated with the full (as opposed to branch-relative) evolution of the initial collapsed state $| Z_{0} \rangle \langle Z_{0} |$ spread beyond the required thresholds. 
Figure ~\ref{fig:MultipleNewton} illustrates this evolution graphically.


From the above analysis, we can see that once decoherence has rendered the initial state of $S$ an incoherent superposition of narrow wave packets, \textit{the stochastic, quasi-classical trajectory of $S$ is virtually certain to be approximately Newtonian over the timescale $T$} on which $(\Delta X(t), \Delta P(t)) < \left(\min\left(2 \delta_{X},l_{V}\right), 2 \delta_{P}\right)$. As a general rule of thumb, $T$ increases with the mass $M$ of the system $S$ and decreases with the size of the Lyapunov exponent $\lambda$ characterizing chaotic effects in $S$'s  classical Hamiltonian \cite{zurek1998decoherence}. Environmental decoherence, while it strongly suppresses coherent spreading in $S$, contributes somewhat to the rate of ensemble spreading in $S$ and so may serve to reduce $T$ (see \cite{schlosshauer2008decoherence}, Ch. 3). For $reduction_{M}$ to hold between the classical model and the stochastic quantum model, these factors must combine to make $T$  greater than or equal to the empirically determined timescale $\tau_{c}$ on which Newtonian models track $S$'s behavior. We can see one role for limits in this analysis by noting that $lim_{M \rightarrow \infty} T = \infty$, which helps us to understand why systems with large mass such as the moon exhibit approximately Newtonian behavior over longer timescales than microscopic systems, such as an alpha particle in a static magnetic field.

\

\

\section{Interpretation Dependence} \label{InterpretationDependence}

In order to extract quasi-classical trajectories from the branching structure generated by the unitary quantum state evolution, we stipulated in Section \ref{SectionFAPPCollapse} that at each moment, some mechanism should select a single branch as the apparently unique, determinate state of affairs that is realized, with probability given by the Born Rule. 
We will now consider the mechanisms suggested by three different interpretations of quantum theory - Everett, de Broglie-Bohm and GRW - that may serve to underwrite this decoherence-compatible collapse process. In doing so, I show explicitly how, in the recovery of determinate, approximately Newtonian trajectories from quantum theory, questions about collapse and ontology can be addressed as a coda - albeit a necessary one - to an interpretation-neutral analysis based on decoherence. In our discussion we will see that the status of ``unrealized" branches varies significantly across these interpretations. An important advantage of this approach to recovering classical behavior is that one need not begin anew in the recovery of classical behavior with each new interpretation that is considered.

\subsection{Everett}

Everettian quantum mechanics posits that the quantum state is a complete description of the state of any closed system and that it always evolves according to Schrodinger's equation. Thus, it takes  collapse to be a merely effective and apparent process. On the Everett interpretation, all branches continue to coexist in the structure of the total quantum state and have an equal claim to physical reality. 
\footnote{For a modern development of the Everett Interpretation grounded in decoherence theory, see Saunders, Kent, Barrett and Wallace's recent volume, \cite{saunders2010many}, and Wallace's \cite{wallace2012emergent}. 
}
As Wallace has observed, from an ontological perspective, recovering the phenomenology of collapse is truly an exercise in interpretation rather than a matter of augmenting or modifying the usual quantum formalism, as it is on the dBB and GRW ``interpretations".

The appearance of determinate, quasi-classical measurement outcomes in Everett's interpretation results from our limited vantage point as observers within the overall structure of the quantum state. 
Within the mathematical formalism, an observer with determinate beliefs about the position and momentum of system $S$ is represented not simply by some subset $O$ of degrees of freedom in $E$, but rather by the state of this subset relative to some particular branch of the overall quantum state. In general, the \textit{total} state of $O$ will be an incoherent superposition of states defined relative to different branches, where each branch-relative state of $O$ corresponds to a different ``copy" of the observer. At a branching event, each such copy splits, like an amoeba, into further copies. While each of these branch-relative observers registers a unique past history for the system $S$, the future of $S$ relative to such an observer is indeterminate since it gives rise to many future copies that register different trajectories for $S$.  

It is also important to distinguish here between the multiplicity of observers that may occur in a single branch and the multiplicity that may occur \textit{across} different branches. In a single branch, there may be multiple observers associated with disjoint subsets $O_{1}, O_{2}, O_{3}, ...$ of degrees of freedom in the environment, and there will be only one copy of each of these observers, and all of these observers should register the same behavior for the system S. Across different branches, there may be multiple copies of a single observer, all of which are associated with the same set of degrees of freedom $O_{i}$ and register different behaviors for $S$. The notion of objectivity that requires different observers to register the same behavior for the system $S$ requires
 consistency among the first type of multiplicity of observers but not among the latter type.

Beyond explaining the emergence
\footnote{Note that I do \textit{not} use the term ``emergence" here in the sense that is sometimes employed in the philosophical literature, which denotes failure of reduction. 
}
of quasi-classicality from the unitary evolution, however, there is also the matter in Everett's interpretation of accounting for the empirical validity of the Born Rule, which enters into our analysis through the probability $|W_{\alpha_{0}...\alpha_{N}}|^{2}$ of effective collapse onto the state $\frac{1}{W_{\alpha_{0}...\alpha_{N}}}  \hat{C}_{\alpha_{0} ... \alpha_{N}} | \Psi_{0} \rangle$. 
Among proponents of the Everett interpretation, there are multiple views concerning the best way to address this issue. The combined work of Deutsch and Wallace has yielded what is widely regarded as one of the most  promising proposals \cite{deutsch1999quantum}, \cite{wallace2007quantum}. More recently, an alternative approach has been suggested by Carroll and Sebens \cite{carroll2014many}. Taking the Everett interpretation on its own terms and allowing provisionally that it does succeed at recovering the quasi-classical Born Rule phenomenology, the account of classical behavior given in Section \ref{NewtonianTrajectories} can be straightforwardly superimposed onto this interpretation by interpreting the collapse of Section \label{SectionFAPPCollapse} as a merely effective process that results from branching of the quantum state.
The main interpretation-specific qualification that must be added to the analysis of Section \ref{FAPPClassical}, beyond an Everettian derivation of the Born Rule, is that the appearance of only one branch being realized is a reflection solely of our limited vantage point as branch-relative observers.



\subsection{de Broglie-Bohm}

According to the de Broglie-Bohm (dBB) interpretation of quantum theory, a complete specification of $SE$'s state consists of both its quantum state and an additional configuration that describes the positions in 3D space of the centers of mass in $S$ and of the particles in the environment $E$ (for details, see \cite{bohm1995undivided}, \cite{holland1995quantum}). The quantum state is postulated to evolve at all times according to Schrodinger's equation while the Bohmian configuration is postulated to evolve according to the deterministic rule $\dot{q_{i}} = \frac{\nabla_{i} S(x,t)}{m_{i}}$, where $S(x,t)$ is the phase of the wave function and $q_{i}$, $m_{i}$ and $\nabla_{i}$ are, respectively, the spatial position, mass and gradient operator associated with the $i^{th}$ degree of freedom in $SE$. It also usually assumed that our ignorance of the Bohmian configuration at some $t=0$ is characterized by $|\Psi(X,y; t=0)|^{2}$, where $X$ is a configuration of $S$ and $y$ a configuration of $E$.
\footnote{Valentini and collaborators have argued that it is possible to derive this assumption from the dynamics of dBB theory rather than simply postulating it. See, for example, \cite{towler2012time} and \cite{valentini2005dynamical}. 
}
The outcomes of measurements and our experience of the world more generally are taken to supervene directly on the Bohmian configuration and not the quantum state; the quantum state determines phenomenology only indirectly through its \textit{dynamical} effects on the Bohmian configuration. Fundamentally, the dynamics governing the evolution of both the quantum state and the Bohmian configuration are deterministic. As in the Everett interpretation, all branches of the quantum state continue to coexist. However, by contrast with the Everett interpretation, the Bohmian configuration serves to single out one of these branches as the only one that is phenomenologically relevant in the sense that it is the only component of the quantum state that exerts any dynamical influence on the Bohmian configuration. Because other branches are dynamically irrelevant to the evolution of the configuration, they can for all practical purposes be ignored even though they continue to exist as part of the theory's ontology. For this type of effective collapse to occur, it is necessary that the different branches be disjoint in the total configuration space of $SE$ and that the process through which they become disjoint be effectively irreversible. Because disjointness in configuration space implies orthogonality, the requirement that different components of the overall state become irreversibly disjoint in the total configuration space is a special form of decoherence. This basic mechanism for effective collapse was first spelled out by Bohm in the second of his 1952 papers; in a sense, Bohm's paper anticipates more modern work on decoherence by several decades, albeit in a highly interpretation-specific context
\cite{bohm1952suggestedii}.

The branching  evolution described in Section \ref{UnitaryBranching}, which is taken to follow from the unitary Schrodinger dynamics, assumes mutual orthogonality of branch states at each time. Following the discussion of the previous paragraph, for dBB theory to successfully underwrite the effective collapse described in equation (\ref{FAPPCollapse2}), it is necessary that the branches at all times be not just orthogonal but mutually disjoint in the total configuration space of $SE$:

\begin{equation} \label{ConfigDec}
\langle \Psi_{0} | \hat{C}_{\alpha'_{0},...,\alpha'_{N}}^{ \dagger}|X,y \rangle \langle X,y | \hat{C}_{\alpha_{0} ...\alpha_{N}}   | \Psi_{0}\rangle \approx 0 \ \forall \ (X,y) \in \mathbb{Q}_{SE} 
\end{equation}

\noindent for $\alpha_{k} \neq \alpha_{k}$ for $0 \leq k \leq N$
\footnote{Hiley and Maroney have called the requirement of disjointness in configuration space ``superorthogonality" \cite{MaroneyDensity}, \cite{bohm1995undivided}. 
}.
Assuming that this requiement holds for all $N$, it follows that the configuration $q_{SE}(N \Delta t)$ of the total system at a given time $N \Delta t$ can only be guided by one branch, which furnishes the effective state of the total system $SE$ at that time. Assuming that $q_{SE}(N \Delta t)$ lies in $\hat{C}_{\alpha_{0} ... \alpha_{N}}   | \Psi_{0}\rangle$, it follows that $q_{SE}((N+1) \Delta t)$ must lie in one of the packets $\hat{C}_{\alpha_{0} ... \alpha_{N} \alpha_{N+1}}   | \Psi_{0}\rangle$. Moreover, conditioned on the assumption that  $q_{SE}(N \Delta t)$ lies in $\hat{C}_{\alpha_{0} ... \alpha_{N}}   | \Psi_{0}\rangle$, 
it can be shown that the probability of $q_{SE}((N+1) \Delta t)$ being in the wave packet $\hat{C}_{\alpha_{0}  ... \alpha_{N} \alpha_{N+1}}   | \Psi_{0}\rangle$  is $\frac{|W_{\alpha_{1}...\alpha_{N} \alpha_{N+1}}|^{2}}{|W_{\alpha_{1}...\alpha_{N}}|^{2}}$.
\footnote{This follows immediately from the analysis of measurement given in \cite{holland1995quantum}, Ch. 8.
}
Note that these probabilities merely reflect our ignorance of the Bohmian configuration $q_{SE}$ rather than any sort of ontological indeterminacy in the system's configuration.  
A more comprehensive presentation of this decoherence-based approach to recovering classical behavior in dBB theory, including a comparison with the popular alternative approach based on the so-called ``quantum potential", is given in \cite{RosalerDBB2015}.

\subsection{GRW}

In GRW theory, collapse of the quantum state is a real - as opposed to merely effective or apparent - dynamical process
\footnote{For an introduction to GRW theory, consult e.g., \cite{bassi2003dynamical}, \cite{ghirardi1986unified}.
}. 
Moreover, by contrast with the Everett and dBB theories, which are deterministic, this collapse is fundamentally stochastic and does not result merely from a restriction on our knowledge of the system's complete state or from a limitation of our vantage point.
However, it is an open question whether, in the systems that concern us here, the fundamentally non-unitary collapse postulated by GRW theory reliably produces one of the branch states $\frac{1}{W_{\alpha_{0}...\alpha_{N}}}  \hat{C}_{\alpha_{0} ... \alpha_{N}} | \Psi_{0} \rangle$ generated by the effective decoherence-compatible collapse prescription presented in Section \ref{SectionFAPPCollapse}. That is, it is an open question whether the GRW collapse mechanism is decoherence-compatible. Whether this is the case will depend in part on whether GRW collapse occurs on timescales shorter or longer than decoherence timescales (for further discussion of this point, see \cite{sep-qm-decoherence} and \cite{tegmark1993apparent}). As Bacciagaluppi has noted, if decoherence timescales are shorter than GRW collapse timescales, then it is possible that the GRW collapses will simply yield one of the branch states defined by decoherence \cite{sep-qm-decoherence}. On the other hand, if decoherence timescales were longer than GRW collapse timescales, this would likely not be the case. If the states produced by GRW collapse are not suitable approximations to the branch states generated by decoherence, there should be a significant, readily testable empirical difference between GRW theory and interpretations that employ decoherence-compatible collapse mechanisms, in which case one ought to be able to rule out one or the other from consideration. 

Assuming that the GRW collapse mechanism is decoherence-compatible, we can understand the effective recipe for collapse employed in Section \ref{SectionFAPPCollapse} as being underwritten on this interpretation by a real, fundamentally stochastic, non-unitary dynamical collapse process. Naively, we might think that the dynamical collapse process of GRW theory described in terms of branch states would be


\begin{flalign} \label{DynamicalCollapseDecoherence}
& \frac{1}{W_{\alpha_{0}...\alpha_{N}}}  \hat{C}_{\alpha_{0} ... \alpha_{N}} | \Psi_{0} \rangle \xrightarrow{dynamical \ collapse} \frac{1}{W_{\alpha_{0}...\alpha_{N} \alpha_{N+1} } } \hat{C}_{\alpha_{0} ... \alpha_{N} \alpha_{N+1}} | \Psi_{0} \rangle \ \ \ \ \ w/ \ prob. \ \ \frac{|W_{\alpha_{0}...\alpha_{N} \alpha_{N+1}}|^{2}}{|W_{\alpha_{0}...\alpha_{N}}|^{2}}
\end{flalign}

\noindent where the renormalization of the quantum state during the collapse process is real rather than merely effective as it is in the Everett and dBB interpretations. However, it is more reasonable to think that the states involved in GRW collapses would be fine-grainings of the branch states that appear in this relation. That is to say, they would be states $\frac{1}{W_{\alpha_{0}...\alpha_{N}}^{\beta_{1}...\beta_{N}}}\hat{C}_{\alpha_{0}...\alpha_{N}} ^{\beta_{0}...\beta_{N}} | \Psi_{0} \rangle$ such that 

\begin{equation}
\hat{C}_{\alpha_{0} ... \alpha_{N}} | \Psi_{0} \rangle = \sum_{\beta_{0}...\beta_{N}} \hat{C}_{\alpha_{0}...\alpha_{N}} ^{\beta_{0}...\beta_{N}} | \Psi_{0} \rangle
\end{equation}

\noindent and  $\langle \Psi_{0}| \hat{C}_{\alpha'_{0}...\alpha'_{N}}^{\dagger, \beta'_{0}...\beta'_{N}} \hat{C}_{\alpha_{0}...\alpha_{N}} ^{\beta_{0}...\beta_{N}} | \Psi_{0} \rangle \approx 0$ if $\alpha'_{i} \neq \alpha_{i}$ or $\beta'_{i} \neq \beta_{i}$ for any $0 \leq i \leq N$. The reasoning behind this claim is as follows. GRW theory's collapse mechanism relies on the fact that although it is extremely unlikely that any given particle will collapse in a given time period, for a collection of $\sim 10^{23}$ such particles, it is extremely likely that one or another of them will undergo collapse and thereby induce collapse on any systems with which it is entangled. In the context of the coupled pair of systems $SE$ considered here, it is virtually certain that the degree of freedom that induces the collapse of the total state of $SE$ will be a particle in the environment $E$ rather than in $S$, since there are many, many more degrees of freedom in $E$ than in $S$. However, the coarse-grained branch states $\frac{1}{W_{\alpha_{1}...\alpha_{N}}}  \hat{C}_{\alpha_{1} ... \alpha_{N}} | \Psi_{0} \rangle$, while narrowly localized with respect to the system $S$, need not be localized with respect to a given degree of freedom in $E$, and thus will not coincide with the states resulting from GRW collapse, which are narrowly localized with respect to any particle that induces the collapse. On the other hand, it is possible that more fine-grained branch states $\frac{1}{W_{\alpha_{1}...\alpha_{N}}^{\beta_{1}...\beta_{N}}}\hat{C}_{\alpha_{1}...\alpha_{N}} ^{\beta_{1}...\beta_{N}} | \Psi_{0} \rangle$ could be found that are localized with respect to all or most particles in the environment and therefore do coincide with the possible products of GRW collapse. From a metaphysical point of view, it would be these more fine-grained branch states rather than the ones we have been discussing that describe the true state of the system $SE$ at each time. The evolution of $SE$'s state would then consist of a rapid-fire alternation between brief periods of unitary evolution (in which the stochastic term of GRW theory is dormant) and non-unitary collapse onto these fine-grained branch states.

On the other hand, given that our primary concern is with the behavior of the system $S$, GRW theory explains why the coarse-grained representation of collapse given in (\ref{DynamicalCollapseDecoherence}) could serve for all practical purposes in describing $S$'s dynamics even though the state evolution that it ascribes to $SE$ is not quite accurate from a metaphysical point of view. All of the fine-grained branch vectors $\frac{1}{W_{\alpha_{1}...\alpha_{N}}^{\beta_{1}...\beta_{N}}}\hat{C}_{\alpha_{1}...\alpha_{N}} ^{\beta_{1}...\beta_{N}} | \Psi_{0} \rangle$ for fixed $(\alpha_{1},...,\alpha_{N})$ ascribe the same branch-relative reduced state $\hat{\rho}_{\alpha_{0} ... \alpha_{N}}$ to $S$ - which is also the state relative to the coarse-grained branch vector - since the fine-graining of these states pertains to degrees of freedom in $E$ and not $S$. If the fine-grained histories are mutually decoherent, the probability of the history $(\alpha_{1}, ..., \alpha_{N})$ on our coarse-grained collapse prescription is equal to the probability that one or another of the fine-grained histories  $(\alpha_{1}, \beta_{1}; ... ;\alpha_{N}, \beta_{N})$ occurs. Since we are mainly concerned with the behavior of $S$, and all of these fine-grained histories agree on $S$'s behavior, it is a matter of indifference which of the possible fine-grained behaviors the environment happens to exhibit, and the transition probabilities and expectation values prescribed by the coarse-grained collapse serve just as well.

\subsection{Summary: Branching and Effective Collapse Across Interpretations}



\begin{figure}[] \centering
\includegraphics[scale=.15]{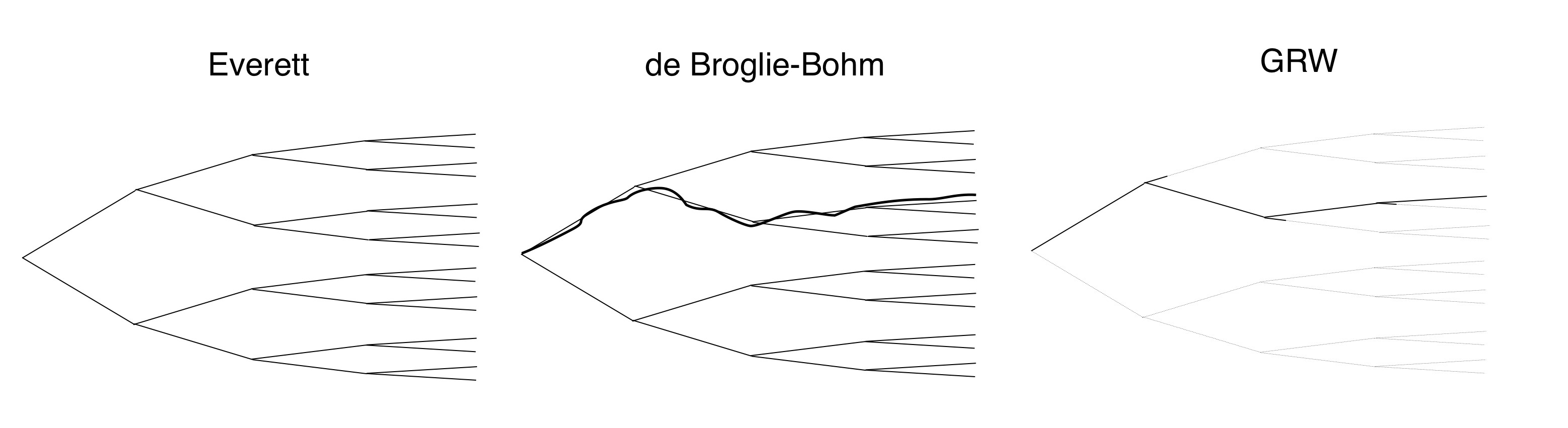}
\caption{\footnotesize The Everett, de Broglie-Bohm and GRW interpretations of quantum mechanics all rely on the branching structure generated through decoherence in order to recover the phenomenology of wave function collapse (assuming, that is, that their collapse mechanisms are decoherence-compatible). According to Everett's interpretation, no modifications to the quantum state or Schrodinger dynamics are necessary to recover the phenomelogy of collapse. According to the dBB interpretation, an additional configuration (whose trajectory is represented by the thick black curve in the middle diagram) singles out one branch as dynamically and phenomenologically relevant; all other branches continue to exist, but are irrelevant to the configuration's trajectory and to the theory's phenomenology. Assuming that the GRW collapse mechanism is decoherence-compatible, the unitary branching structure represents the set of all possible stochastic state evolutions on GRW theory; while only one such evolution occurs (bold line in the right-hand figure), the unrealized branches (represented by the dotted lines in the right-hand figure) may be interpreted as evolutions that could have occurred on this evolution. 
}
\label{fig:BranchingInterpretations}
\end{figure}

We have seen that different interpretations employ widely varying mechanisms for explaining how and in precisely what sense just a single branch in the unitary quantum state evolution comes to characterize a system's behavior, and for explaining why it does so in accordance with the Born Rule. 
Because the collapse mechanisms of different interpretations all fill this common functional role in the explanation of classical behavior, it is possible to ``black box" the details of this mechanism and to gain a reasonably precise quantum mechanical picture of classical behavior 
 \textit{without committing} to any particular interpretation, as we did throughout Section \ref{FAPPClassical}. For obvious reasons, this point should be of special interest to interpretational agnostics. Figure ~\ref{fig:BranchingInterpretations} illustrates schematically the different ways in which the three interpretations considered in this section rely on the unitary branching structure in recovering the phenomenology of collapse. 

It should also be noted that the precise restrictions on the quantum state required for collapse or effective collapse to occur vary across interpretations. Whereas simple orthogonality of distinct branch states suffices for this purpose on the Everett interpretation, a stronger, more restrictive form of decoherence, associated with the requirement of disjointness of different branches in configuration space, must hold before the Bohmian collapse mechanism takes effect. Most likely, some such stronger decoherence condition is also required to ensure that GRW collapses produce branch states, although the detailed form of this requirement requires further theoretical analysis. Since the conditions on the quantum state required for effective collapse on dBB theory, and probably GRW theory as well, are more restrictive than those required by Everett's theory, the set of states satisfying the stronger condition must be smaller than, and strictly contained in, the set of states satisfying the weaker Everettian condition. This observation suggests the in-principle possibility that the Everettian condition for effective collapse might be satisfied on shorter timescales than the stronger conditions imposed by the dBB and GRW interpretations 
\footnote{This suggestion assumes that the Everettian condition for effective collapse does indeed coincide with simple orthogonality of branch states - as is often supposed - rather than with one of the stronger conditions associated with dBB or GRW theory.
}
That is, it is conceivable that a system might have to pass through a region of states that satisfy the weak but not the strong condition before reaching those states that satisfy the strong condition.  Whether this turns out to be the case, and if so, what the relevant difference of timescales turns out to be, and whether there are any empirically testable consequences to this difference, will depend in part on the detailed dynamics that govern the state's evolution. I leave this as a topic for further investigation.

\section{Concerns about the Decoherence-Based Approach to Recovering Classical Behavior} \label{DecoherenceConcerns}

For the sake of transparency, I wish to briefly highlight several important concerns about the conventional assumptions of decoherence theory and of our analysis in particular. However, I emphasize that my purpose in this article is not to defend decoherence theory from its critics, but rather to clarify the picture of classical behavior that decoherence theory makes available if one provisionally accepts its basic assumptions.

\subsection{Justifying the System/Environment Split}

The conventional setting for decoherence-based analyses of classical behavior described in Section \ref{Setup} takes for granted a privileged splitting of the universe into subsystems $S$ and $E$, where $S$ consists of some composite degrees of freedom such as the center of mass of a macroscopic body and $E$ represents all other degrees of freedom, including the internal degrees of freedom of the body itself. However, at a more basic level of description, all degrees of freedom in the system are microscopic and the quantum model simply describes a large number $N$ of atomic or subatomic particles. From this microscopic perspective, the grounds for separating degrees of freedom into those that constitute the ``body" in question and those that do not, and beyond this, into the center-of-mass and ``residual" degrees of freedom within the body, is less transparent. A full quantum mechanical explanation of classical behavior requires that we be able to motivate the system-environment splitting, and to derive the dynamics used to describe it, from this more basic microscopic perspective.  
Presumably, the separation is grounded partly in contingent features of the overall state of the closed system - for example, a baseball at sufficiently high temperature (a property of the baseball's state) will cease to be a distinct, cohesive object, and the system-environment split between the baseball's center of mass and all other degrees of freedom that is employed at lower temperatures will cease to be physically salient. The separation may also be grounded partly in the form of the microscopic Hamiltonian, which may play a role in motivating natural separations among different subsets of degrees of freedom. The problem as to what, on a fundamental microscopic level, differentiates ``system" from ``environment"  has been acknowledged as a serious puzzle for the decoherence program by Zurek, and has also been discussed in the work of Lombardi \textit{et al}, Schlosshauer, Dugic \textit{et al}, Fields, and Zanardi, among others \cite{ZurekProcRoySoc}, \cite{schlosshauer2005decoherence}, \cite{lombardi2012problem}, \cite{dugic2012parallel}, \cite{fields2014ollivier}, \cite{zanardi2001virtual}.

\subsection{Concerns about Irreversibility}
Decoherence theory rests on the assumption of a certain irreversibility in the evolution of the quantum state: initially coherent superpositions of a system become incoherent superpositions, but not the reverse; the branching structure generated by decoherence exhibits branching toward the future but not the past. On the other hand, the fundamental dynamics from which this irreversible behavior is purported to arise - the unitary Schrodinger dynamics - is time reversible, so that any series of states that solves the equations of motion also solves them when time-reversed. Much as in the foundations of classical statistical mechanics, there exists a \textit{prima facie} conflict between a certain set of irreversible phenomena and the time reversibility of the dynamics that purportedly underwrite it.  
\footnote{For a discussion of this issue from the perspective of Everett's interpretation, see \cite{wallace2012emergent}, Ch. 9.
}

A recent critique by Kastner of the decoherence program, and of decoherence-based formulations of Everett's interpretation, calls attention to this problem \cite{kastner2014einselection}. While authors like Zurek, Saunders and Wallace claim that decoherence and branch-relative quasi-classicality can be derived as a generic consequence of the unitary Schrodinger evolution for macroscopic systems, Kastner argues that existing arguments in support of this claim are circular. The circularity, she claims, lies in the fact that quasi-classicality is effectively smuggled into the premises of the argument through a choice of initial conditions that are themselves quasi-classical. According to Kastner, this strongly parallels the circularity of Boltzmann's molecular chaos assumption in his proof of the H-Theorem (which is thought to smuggle in the irreversibility that Boltzmann purported to derive). As she writes, ``Everettian decoherentists have effectively assumed what they are trying to prove: macroscopic classicality only `emerges' in this picture because a classical, non-quantum-correlated environment was illegitimately put in by hand from the beginning. Without that unjustified presupposition, there would be no vanishing of the off-diagonal terms and therefore no apparent diagonalization of the system's reduced density matrix that could support even an approximate, `FAPP' mixed state interpretation." A full response to Kastner's worry, and to the broader question of how to reconcile the irreversibility of decoherence with the reversibility of the Schrodinger dynamics, is beyond the scope of this article. However, one line of response to Kastner's more specific charge of circularity would be to argue that decoherence does not occur only for the special choice of initial conditions that Kastner claims to lie at the root of the circularity, but also for a much more general class of states in which the system and environment may already be entangled and where the system's reduced density matrix may \textit{already} be approximately diagonal in position. The relevant notion of decoherence here is not restricted to the vanishing of off-diagonal elements of a system's reduced density matrix but extends more generally to branching of the quantum state, which can also occur for states that are already entangled. 

\subsection{Concerns about Chaos}

One of the most pervasive sources of skepticism concerning the reduction of classical to quantum theory has been the claim that quantum theory cannot recover the exponential divergence between trajectories characteristic of classically chaotic behavior. While classically chaotic evolution may cause two phase space points $Z$ and $Z'$ that are initially nearby to rapidly move apart, unitarity of the Schrodinger evolution (which preserves inner products between states over time) precludes the same sort of  divergence between the corresponding wave packets $| Z \rangle$ and $| Z' \rangle$. Decoherence theory offers a possible route around this worry through the fact that the effective subsystem dynamics of an open system $S$ need not be unitary, and are therefore potentially consistent with an exponential divergence between the wave packets $| Z \rangle$ and $| Z' \rangle$ in $S$'s Hilbert space. Zurek and Paz  have offered a more detailed decoherence-based analysis of classically chaotic behavior, and cite the success of decoherence in resolving tensions between chaos and unitarity as a major point in its favor. On the other hand, Wiebe and Ballentine have argued that classically chaotic behavior can be accounted for in quantum theory without invoking environmental decoherence \cite{wiebe2005quantum}. Further references and a clear summary of this debate can be found in \cite{bokulich2008reexamining}. 

\subsection{Other Assumptions}

The strategy for quantum-classical reduction outlined in Section \ref{FAPPClassical} rests on a number of unproven assumptions apart from those already discussed. First, we have seen that for this strategy to succeed, the time scale $T$ for ensemble spreading of an initial coherent state of $S$ must be no less than the timescale $\tau_{c}$ on which the classical model tracks the behavior of the system in question. I have given no proof that this is the case for real classical systems, simply offering a few general remarks about the factors that affect this rate of spreading. Second, a full reduction of classical to quantum theory requires a derivation of the master equation (\ref{DecoherenceMaster}) from the full Schrodinger dynamics for $SE$, whose details I have left unspecified. While such derivations have been given for models of quantum Brownian motion, 
\footnote{See, for example \cite{schlosshauer2008decoherence}.
}
the full Schrodinger dynamics of realistic quantum systems is a great deal more complicated than this model. A more complete account of classical behavior would explain the validity of the master equation (\ref{DecoherenceMaster}) in the context of more realistic models.

\section{Conclusion}  \label{Conclusion}

The general strategy for quantum-classical reduction consolidated here rests on three main pillars: 1) decoherence, which generates a branching structure for the quantum state, 2) a decoherence-compatible mechanism for collapse, which selects one of these branches as the state of affairs that is realized is some sense, with probability given by the Born Rule, 3) an appropriate form of Ehrenfest's Theorem, which ensures that localized branch-relative trajectories of the system in question are approximately classical over timescales where ensemble spreading in the system is sufficiently minimal. 
This basic strategy can likely be exported to other reductions between classical and quantum models: for example, between models of classical field theory and quantum field theory and of classical gravity and quantum gravity. Both quantum field theory and quantum gravity incorporate the basic linear and unitarity that one finds in non-relativistic quantum mechanics. Because decoherence and Ehrenfest's Theorem are products of a unitary evolution,  
\footnote{The open-systems form of Ehrenfest's Theorem follows indirectly from the unitary evolution for the system $SE$ via the reduced dynamics of $S$.
} 
it is conceivable that one could employ a strategy based on these same three pillars to recover classical field theoretic behavior from quantum field theory and classical gravitational behavior from some quantum theory of gravity. As things stand, this strategy remains underdeveloped in both cases, though some progress has been made in the former. The subject of decoherence and the emergence of pointer states in quantum field theory has been studied by Anglin and Zurek, Kiefer, and Breuer and Petruccione, although the area remains far less developed than the subject of decoherence in non-relativistic quantum mechanics \cite{anglin1996decoherence}, \cite{kiefer1995irreversibility}, \cite{breuer2000radiation}. 
\footnote{It is also explored at length in my doctoral thesis, \cite{RosalerThesis} (copy available on request). 
}
In the realm of quantum gravity (e.g., models of string theory or loop quantum gravity), the subject of decoherence is far less understood. Where questions of interpretation are concerned, 
the Everett interpretation in some sense extends trivially to these realms, although its empirical adequacy in these contexts depends on whether decoherence singles out an appropriate set of quasi-classical states in relevant contexts. There has been some progress in extending the dBB and GRW interpretations to the realms of quantum field theory, relativistic quantum mechanics and even quantum gravity, although the technical issues involved are formidable and the empirical viability of these models remains unclear 
\footnote{See, for example, \cite{StruyvePWQFT}, \cite{struyve2007field}, \cite{struyve2006new}, and \cite{tumulka2006relativistic}, \cite{nikolic2007bohmian}. 
}.
Clearly, there is much work be to done, both in elaborating this three-step, interpretation-neutral strategy for recovering classical behavior in the familiar context of NRQM and in extending it to the context of more general quantum theories. 

\vspace{5mm}

\noindent \textbf{Acknowledgments:} This work was supported by the University of Pittsburgh's Center for Philosophy of Science. Thanks to John Norton, Jos Uffink, Michel Janssen and Geoffrey Hellman for their detailed feedback, and to Leah Henderson, Aris Arageorgis, Slobodan Perovic, Marco Giovanelli, Oliver Pooley and audiences in Oxford, Pittsburgh and Minnesota for helpful commentary. Thanks to David Wallace, Simon Saunders, Christopher Timpson and Jeremy Butterfield for their detailed feedback on early material that forms the basis for much of this work. 

\def\bibsection{\section*{References}}
\scriptsize
\bibliographystyle{plain}
\bibliography{DPhilRefs}

\end{document}